\documentclass[useAMS,usenatbib]{mn2e}

\usepackage{graphicx}
\usepackage{lscape}
\usepackage{float}
\usepackage{epsfig}
\usepackage{multirow}

\usepackage{bigdelim}
\usepackage{bigstrut}

\def\aap{Astronomy \& Astrophysics}
\def\apj{The Astrophysical Journal}
\def\mnras{Monthly Notices of the Royal Astronomical Society}

\title[The SLUGGS survey: chromo-dynamical modelling of the lenticular galaxy NGC\,1023]{The SLUGGS survey: chromo-dynamical modelling of the lenticular galaxy NGC\,1023}
\author[Cortesi, Chies-Santos, Pota, Foster et al.]
{Arianna Cortesi$^{1}$, 
Ana L. Chies-Santos$^{1,2,3}$, 
Vincenzo Pota$^{4}$,
Caroline Foster$^{5}$,
\newauthor
Lodovico Coccato$^{8}$,
Claudia Mendes de Oliveira$^{1}$,
Duncan A. Forbes$^{6}$,
\newauthor
Michael M. Merrifield$^{3}$,
Steven P. Bamford$^{3}$,
Aaron J. Romanowsky$^{4,7}$,
\newauthor
Jean P. Brodie$^{4}$,
Sreeja S. Kartha$^{6}$,
Adebusola B. Alabi$^{6}$,
\newauthor
Robert N. Proctor$^{1}$,
Andres Almeida$^{9}$
\\
$^{1}$Departamento de Astronomia, Instituto de Astronomia, Geofisica e Ciencias Atmosfericas da USP, \\
Cidade Universitaria, CEP:05508900, Sao Paulo, SP, Brazil\\
$^{2}$Departamento de Astronomia, Instituto de F\'isica, Universidade Federal do Rio Grande do Sul, Porto Alegre, R.S, Brazil\\
$^{3}$School of Physics and Astronomy, The University of Nottingham, University Park, Nottingham, NG7 2RD, UK\\
$^{4}$University of California Observatories, 1156 High Street, Santa Cruz, CA 95064, USA\\
$^{5}$Australian Astronomical Observatory, PO Box 915, North Ryde, NSW 1670, Australia\\
$^{6}$Centre for Astrophysics \& Supercomputing, Swinburne University, Hawthorn, VIC 3122, Australia\\
$^{7}$Department of Physics and Astronomy, San Jos\'e State University, One Washington Square, San Jose, CA 95192, USA\\
$^{8}$European Southern Observatory, Karl-Schwarzschild-Straße 2, 85748 Garching bei Muenchen, Germany\\
$^{9}$Department of Physics and Astronomy, Universidad de La Serena, Cisternas, 1200 La Serena, Chile\\
}

\begin{document}

\date{Accepted  Month xx. Received  Month xx;}

\pagerange{\pageref{firstpage}--\pageref{lastpage}} \pubyear{2014}

\maketitle

\label{firstpage}

\begin{abstract}
Globular clusters (GCs) can be considered discrete, long-lived, dynamical tracers that retain crucial information about  the assembly history of their parent galaxy.
 In this paper, we present a new catalogue of GC velocities and colours for the lenticular galaxy NGC\,1023, we study their kinematics and spatial distribution, in comparison with the underlying stellar kinematics and surface brightness profile, and we test a new method for studying GC properties. Specifically, we decompose the galaxy light into its spheroid (assumed to represent the bulge + halo components) and disk components and use it to assign to each GC a  probability of belonging to one of the two components. 
Then we model the galaxy kinematics, assuming a disk and spheroidal component, using planetary nebulae (PNe) and integrated stellar light.
We use this kinematic model and the  probability previously obtained from the photometry to recalculate for each GC its likelihood of being associated with the disk, the spheroid, or neither.
We find that the reddest  GCs are likely to be associated with the disk, as found for faint fuzzies  in this same galaxy, suggesting that the disk of this S0 galaxy originated at $z\simeq2$. The majority of blue GCs are found likely to be  associated with the spheroidal (hot) component.  The method also allows us to identify objects that are unlikely to be in equilibrium with the system. In NGC\,1023 some of the rejected GCs form a  substructure in phase space that is  connected with NGC\,1023  companion galaxy. 

\end{abstract}

\begin{keywords}
galaxies: elliptical and lenticular, cD, galaxies: individual: NGC\, 1023, galaxies: kinematics and dynamics.
\end{keywords}

\section{Introduction}
Understanding the origin of lenticular galaxies and how closely related they are to spiral galaxies is still a challenge  for contemporary astrophysics. A wealth of evolutionary evidence links S0 galaxies to spirals. The morphology-density relation (\citealt{dressler97}) shows that while the number of spiral galaxies decreases  towards the centre of a cluster, the number of S0 galaxies increases. It has also been shown that S0 galaxies are more common at the present epoch than at higher redshifts, while the inverse is true for spiral galaxies (\citealt{dressler83,desai07}). 
The mechanism responsible for transforming a spiral  into an S0 galaxy has to stop star formation in the disk and enhance the spheroidal component (\citealt{dressler83}). Interestingly, S0 galaxies are found in all environments, from high density clusters to the field, allowing for a variety of evolutionary paths (gas stripping, strangulation, harassment, minor mergers, pestering, secular evolution) and raising the question of whether S0 galaxies are a unique class, or a collection of objects whose formation mechanisms are environment dependent (\citealt{Gunn,Salamanca,Quilis,Kronberger,Byrd,bournaud05}). Recent work, moreover, suggests the possibility of creating S0s through major mergers under  specific initial conditions (\citealt{Borlaff2014,naab14}).

The past history of a galaxy is imprinted in its faint outer regions, where we can find signs of minor mergers, traces of accretion, or of a gentle passive fading. In general, it is difficult to study the kinematics and the stellar populations  of early-type galaxies at large radii due to the absence of an undisturbed HI disk and the low surface brightness of the stellar component at large radii. Recently, this last issue has been solved by using discrete kinematic tracers, which are detectable at large radii. Large datasets are becoming available from globular clusters  (\citealt{pota13}, \citealt{Brodie14}), planetary nebulae (PNe) (see e.g. \citealt{c09}, \citealt{cortesi13a}) and integrated stellar light (see eg. \citealt{norris08}, \citealt{foster11}, \citealt{Brodie14}).   
Moreover, GCs provide us with a unique tool for understanding the chemodynamical properties of their host galaxies. They tend to be old ($\ga$\,10\,Gyrs) and are the closest examples we have to simple stellar populations (\citealt{bs06}). GCs are ubiquitous across a range of galaxy types and luminosities, and can be  studied up to distances of $\sim$\,100Mpc (see \citealt{harris10}) and beyond (\citealt{alamo-mart13}).
They form a bimodal distribution in optical-colours that is often interpreted as evidence for metallicity bimodality (\citealt{bs06, brodie12, Cantiello}; for an alternative view see  \citealt{yoon06}, \citealt{cs12a}). 

 In general, blue GCs tend to reside in the galaxy halo, while red globular clusters are associated with the metal rich stellar population,  i.e. the overall stellar population in elliptical galaxies (\citealt{pota13}) and the bulge or the thick disk of spiral galaxies (\citealt{minniti95}, \citealt{forbes01}).
Studying the properties of the GC sub-populations in lenticular galaxies can help constrain S0     formation histories. If S0 galaxies are gently quenched spiral galaxies, we expect their GC sub-populations to have the same properties as in spiral galaxies, only their specific frequency would be higher, since the total number of GCs would be the same in S0s and spirals, while spiral galaxies would be brighter (\citealt{salamanca06}). In contrast, violent events, such as mergers  or the accretion of small companions,  would alter the initial distribution of GC in the galaxies (\citealt{k12}). \cite{forbes12} studied the red GC system in the lenticular galaxy NGC\,2768, finding that the red GCs share the same properties as PNe belonging to the spheroidal component of the galaxy. This suggests that, in this S0, the red GCs behave as in  spiral galaxies.  
However this work is based on only  a colour selection of the GC sub-populations. 


In an attempt to study the GC sub-populations in S0 galaxies to investigate their formation histories, we present a new method  which uses the galaxy photometry and kinematics, as recovered from PNe and integrated stellar light, to obtain the likelihood for each GC of belonging to the spheroid or the disk of a galaxy. This method ensures a physically meaningful  separation, based on the kinematics rather than the colour, and allows us to trace streams and substructures. Subsequently, we construct a chromo-dynamical model of the galaxy, where we correlate  the GC probability of belonging to one of the galaxy components with its colour. 
Our final goal is to understand the build up of the blue and the red GC sub-populations in light of the hierarchical merging framework of galaxy formation and shed light on the possibility of using GCs as probes of galaxy formation.
In particular in this paper we study the lenticular galaxy NGC\,1023. A number of studies have been carried out on this galaxy (\citealt{LB00}, \citealt{cortesi11}, \citealt{cs13}, \citealt{forbes14}; and many others) and it is an ideal case for testing this method. A new catalog of GC positions and velocities has been obtained for this work by the SLUGGS team (see below) and  is published in this paper.


The paper is structured as follows. In Section 2 we present the data we employ for our analysis. The method is presented in Section 3 and the results are presented in Section 4. In Section 5 we discuss our findings in the context of galaxy evolution. Summary and conclusion are presented in Section 6.

\section[]{Data}

In this paper we study three different types of kinematics tracers: PNe, GCs and integrated stellar light. In particular, we test a new method where we use the galaxy kinematics as recovered using PNe and stars to define a likelihood for a given GC to belong to the disk or the spheroid of the galaxy. 
All the data have been published previously with the exception of the GCs that are published here. In the present section we briefly describe the detection techniques and data reduction analysis of the three different datasets.

\begin{figure*}
\includegraphics[scale=0.6]{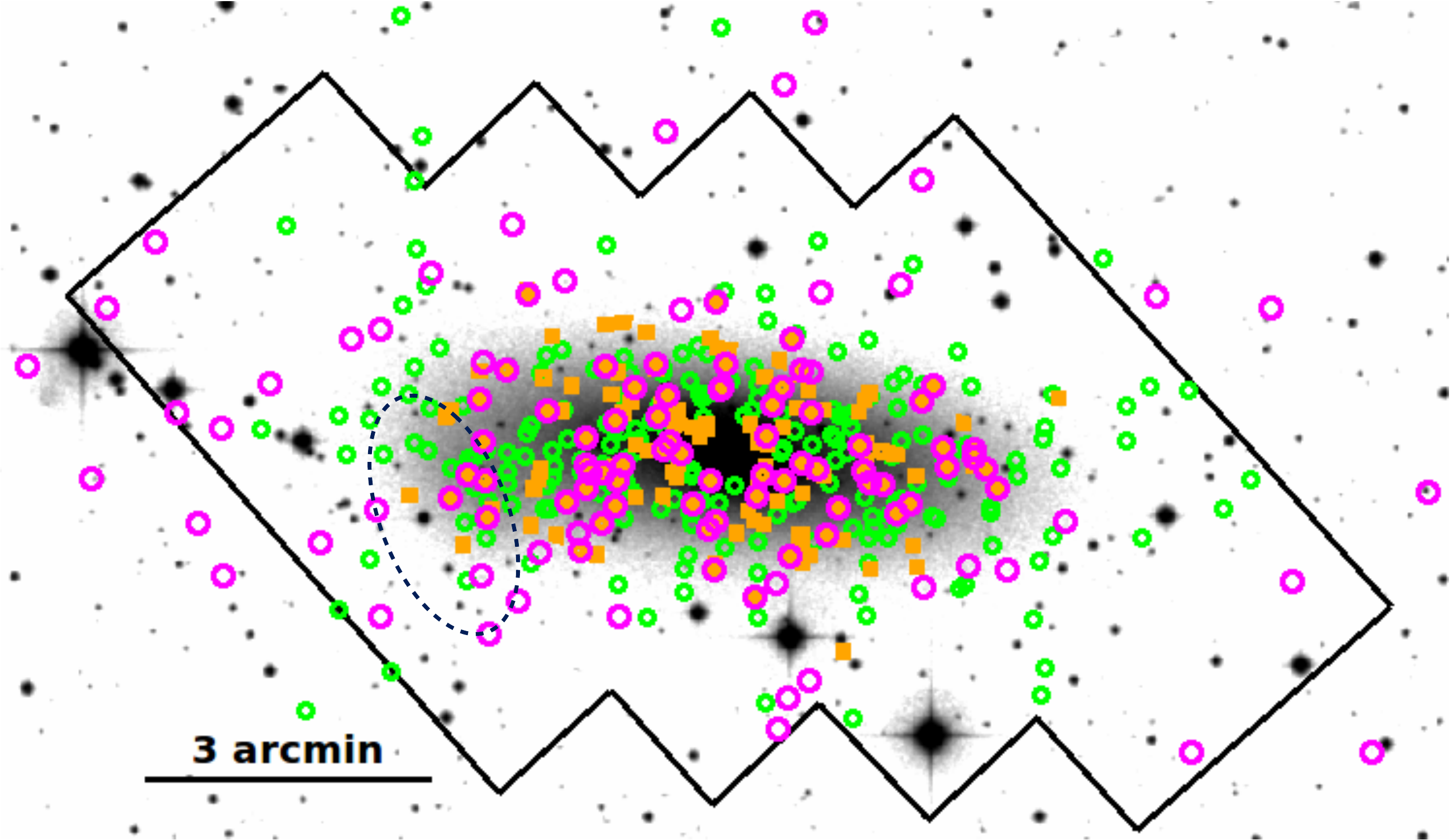}
\caption{Spatial distribution of the spectroscopically confirmed tracers around NGC~1023. The image is from DSS ( about $20 \times 9$ arcmin$^2$ across), with the HST field-of-view outlined in black. Symbols are: diffuse starlight (orange boxes), PNe (small, green, open circles), GCs (large, magenta, open circles).  The location of the companion galaxy is shown as a dotted ellipse,  obtained by ellipse fitting of  a B-band image of the galaxy,  the companion galaxy has bluer colours than the main galaxy \citep{Nor08}. North is up, East is left.}
\label{fig:Figure1}
\end{figure*}

Fig.\,\ref{fig:Figure1} shows the spatial distribution of PNe (small, green, open circles), GCs (large, magenta circles) and integrated stellar light (filled, orange squares) over a DSS image of NGC\,1023, with the HST field-of-view outlined in black. Note the large radial extent that such tracers cover, especially PNe and GCs that reach out to  $\sim 7.46$ disk scale length, $R_{d}$, with $R_{d} \simeq 67$ arcsecs (\citealt{cortesi13b}). East of the galaxy it is possible to see an excess of light at the location of  the companion galaxy NGC\,1023A. We have adopted a distance for the galaxy of 11.1 Mpc  \citep{Brodie14}.

\subsection[GC]{Globular Clusters}

GC data were acquired as part of the SAGES Legacy Unifying Globulars and GalaxieS (SLUGGS) Survey\footnote{http://sluggs.swin.edu.au/Start.html} \citep{Brodie14}. GC candidates were identified in HST/ACS imaging in $g$ and $z$ filters, as described in \citet{forbes14}. The latter was supplemented with archival CFHT data in $g$ and $i$ from \citet{Kartha} to target GC candidates outside the HST footprint. For both datasets, GC candidates were selected based on their location in the colour-magnitude diagram and, for HST imaging only, based on their partially resolved half-light radius $R_{e}$. We adopt the classification of \cite{norris14}, for which GCs are fainter than $-10$ mag in g-band and ultra compact dwarfs (UCDs) are brighter than $-10$ mag. The radius definitions to divide GCs from faint fuzzies (FFs, \citealt{LB00}) are somewhat arbitrary, in this work we follow the definition of \citealt{forbes14}, for which FFs have $R_{e} > 7$ pc and GCs have $R_{e} < 7$ pc. The total number of GCs recovered using only photometry ({\it photometric GCs}) is 360.

\begin{figure*}
\includegraphics[width=1\textwidth]{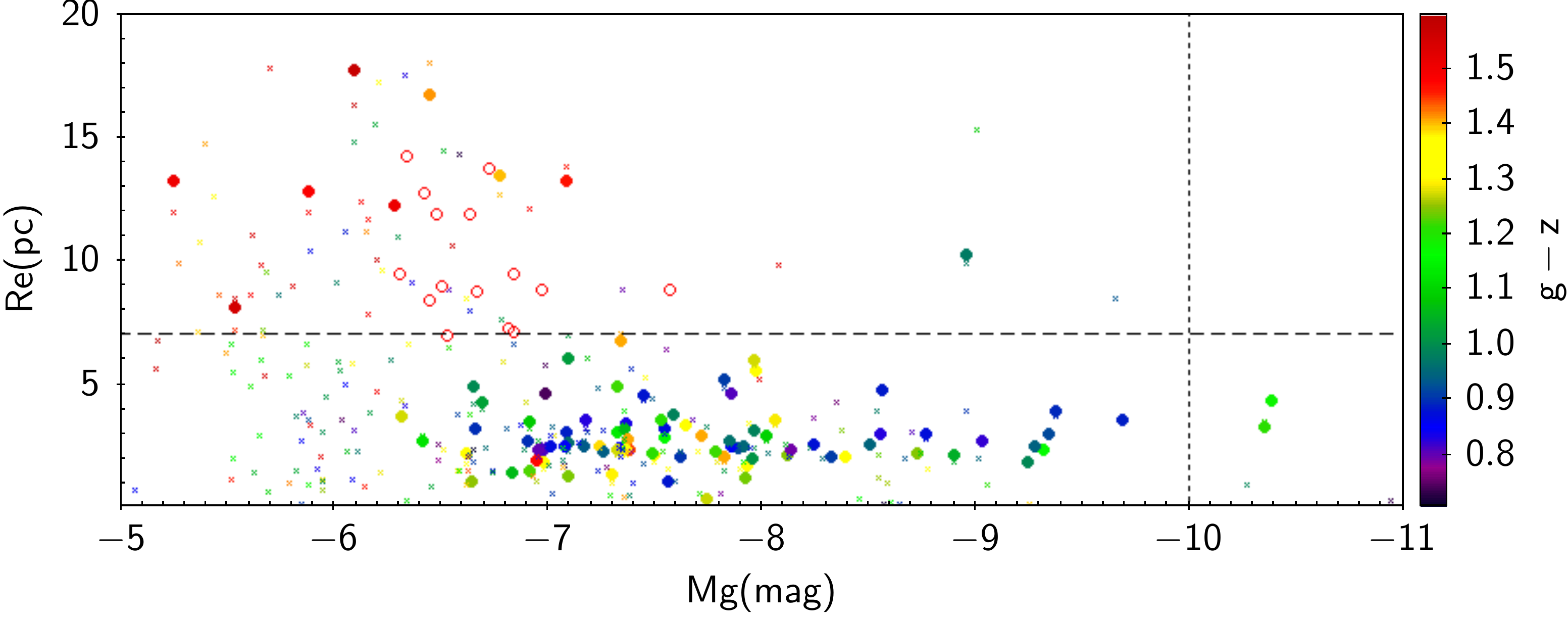}
\caption{Classification of the spectroscopically confirmed objects. Effective radius vs absolute magnitude diagram for the objects published in this paper (big filled circles), colour coded according to their $g-z$ colour and for the FFs from the literature (\citealt{LB00}). For comparison we plot the objects from the photometric catalog, as crosses, used to select the spectroscopic candidates, also colour coded according to their $g-z$ colour. The absolute magnitude was obtained by subtracting the apparent magnitude from the distance modulus value found in the  NED database. The horizontal line at $R_{e}=7$ pc marks the division between GCs and FFs, and the vertical line at $M_{g}=-10$ mag, shows the objects identified as UCDs.}
\label{fig:Figure1bis}
\end{figure*}

Spectroscopic follow-up of a subsample of GC candidates was performed with the multi-object spectrograph Keck/DEIMOS. A total of four DEIMOS masks were observed, see Table \ref{tab:masks}. Two masks were aligned along the major axis of the galaxy, and two along the minor axis. Priority was given to bright and partially-resolved GCs in HST, whereas remaining slits were assigned to un-resolved objects outside the HST field-of-view. The reader should see  \cite{pota13} for a detailed description of the data reduction, while we summarise here the main points of the data analysis. We measure the redshift of the Calcium Triplet (CaT) lines at 8498\AA\ , 8542\AA\ and 8662\AA\. This was performed by cross-correlating eleven template spectra of Galactic stars spanning a wide range of metallicities, with the actual GC spectrum.
Figure \ref{fig:Figure1} shows that most of the confirmed GCs turned out to be inside the HST field-of-view. This is because the GC system of NGC~1023 is $\sim 6$ arcmin in diameter and it is elongated along the photometric major axis \citep{Kartha} as confirmed by the paucity of GCs along the galaxy minor axis. 

\begin{table}
\caption[Summary of the observation details for NGC\~1023.]{Summery of the observation details for NGC\~1023. \label{tab:masks}}
\centering  
\begin{tabular}{c c c c c c}
\hline\hline
Mask & Date &  Exposure time  \\
\hline
$N1023\_1$  & $2011-11-31$ & $1800*1, 1799*1, 1799*1$\\
$N1023\_2$  & $2011-11-31$ & $1800*2, 1799*1$ \\
$N1023\_3$  & $2012-01-15$ & $1800*4, 1639*1$ \\
$N1023\_4$  & $2013-01-16$ & $1800*2, 1320*1$ \\
$N1023\_3$  & $2013-09-29$ & $1800*4$ \\
\hline
\noindent
\end{tabular}
\end{table}

\begin{figure}
\includegraphics[width=0.45\textwidth]{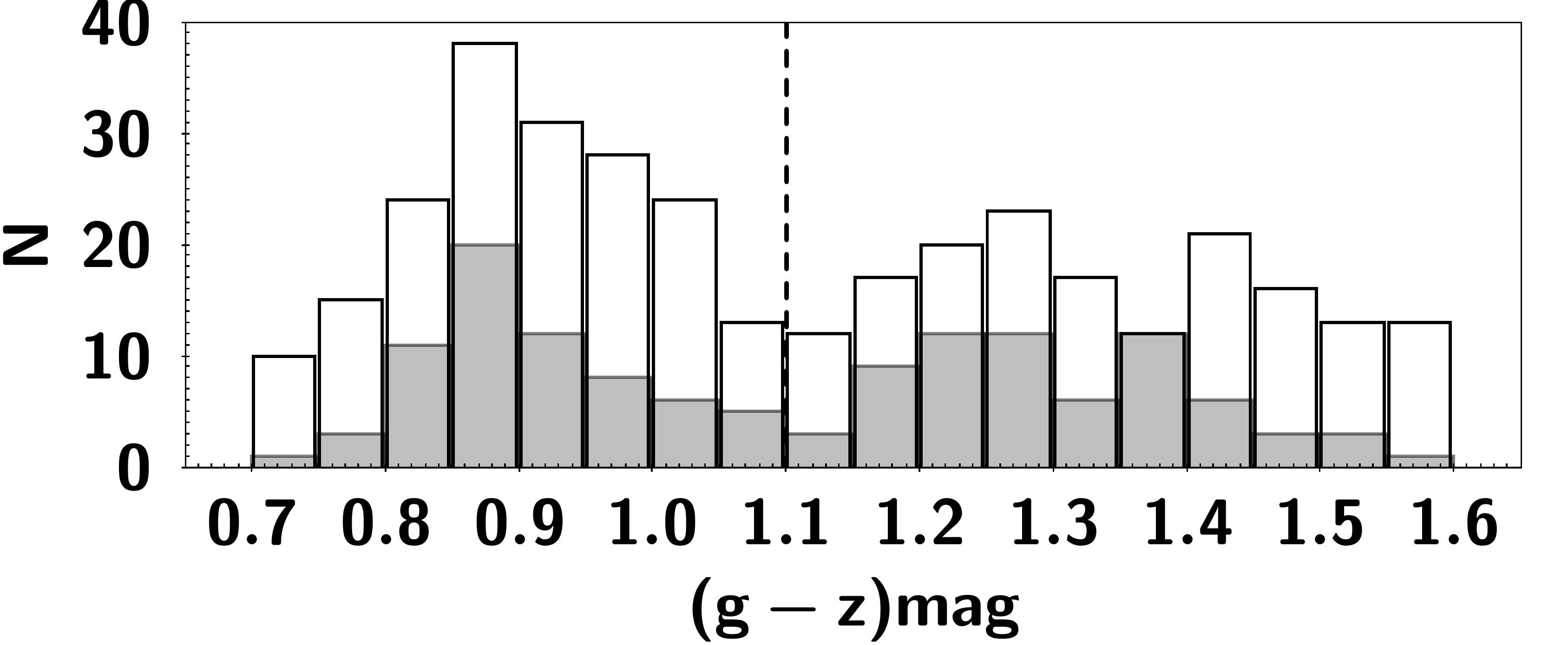}
\caption{GC colour bimodality. Histogram showing the distribution of the (g-z) colour of the GC population, open histogram, and of the spectroscopically confirmed sample, filled histogram. The  dashed line shows the separation between blue and red GCs.}
\label{fig:Figure1tris}
\end{figure}

We adopt the HST catalog as our primary photometric catalog. We convert the CFHT $(g-i)$ colours into HST $(g-z)$ colours by matching the objects in common between the two datasets. We find $(g-z) = 0.39 + 0.59 (g - i)$ and we apply this conversion to the objects with ground-based photometry only.
The $(g-z)$ GC colour distribution is bimodal, see Figure \ref{fig:Figure1tris}, with a colour separation between blue and red GCs set at $(g-z)=1.1$ mag \citep{forbes14}. 

The spectra for 115 candidates belonging to the NGC 1023 systems have been obtained and the velocities have been confirmed by two independent observers. 
For 96 of the 115 detections, which fall in the fields observed with HST, we have measured the effective radius, $R_{e}$. Following \cite{forbes14},  we identify as FFs objects with $R_{e}>7$ pc and  as UCDs objects with $M_{g} < -10$ mag, see Figure \ref{fig:Figure1bis}. In Figure \ref{fig:Figure1bis}, we also show the location of the FFs observed in this galaxy by \cite{LB00}, which will be used in the following sections to complete our FFs catalog. The remaining 19 objects with CFHT photometry are also classified as GCs, since they have a magnitude  $M_{g} > - 10$ mag in g band and lie in the outskirts of the galaxy, while FFs are generally found close to the galaxy centre. In total we have 104 GCs, 63 blue and 41 red ({\it spectroscopic GCs}).
Following \cite{Nor08} we have assigned 6 GCs  to the companion galaxy using the luminosity ratio between the two galaxies and requiring that the radial velocity of the object has a maximum difference with respect to  the systemic velocity of NGC\,1023A of $|\Delta V|\le100$ km/s. The companion galaxy GCs are 3 red and 3 blue.

\subsection[]{Stellar data}

The integrated stellar light   individual data points were presented in \citealt{arnold13} and the integrated stellar light data are taken from  the kinemetry fits presented in Foster et al. (2015, submitted), they were extracted using the SKiMS technique (\citealt{foster13}).  

The method used to extract integrated kinematic information of the underlying galaxy starlight from a multi-slit spectrograph was first developed by \cite{norris08} and  \cite{proctor09}.

 We use the individual data points to assign a probability of belonging to the companion galaxy,  following the same procedure as for PNe and GCs. We find  that 90 \% of the observed stellar light is associated with the main galaxy.
We use the binned kinemetry fits to compare the stellar kinematics with the kinematics of PNe and GCs, see Figure \ref{fig:Figure3}.  

\subsection[]{Planetary nebulae}

The PNe catalog was obtained using the Planetary Nebulae Spectrograph (PNS. \citealt{douglas02}). It was published in \cite{Nor08} and information on the data reduction and completeness correction can be found in \cite{cortesi13a}: where 203 PNe were detected, 20 of which are associated with the companion galaxy N1023A. The PNe  trace the galaxy light profile well, once corrected for incompleteness in the centre. When fitting with a tilted disk model the PNe present a strange velocity profile, with decreasing rotation velocity and increasing values of the tangential components of the random motions (\citealt{Nor08,cortesi11}). This was explained as a contamination  of  the disk kinematics by spheroid PNe. When this effect is properly accounted for, by assigning a probability to each PN of belonging to the spheroid or the disk,  the derived kinematic presents a spiral-like behaviour, with a flat rotation curve and decreasing dispersion velocity in the disk. 

\section{Analysis}

The colours of GCs are believed to be connected to the 
galaxy formation history. One scenario (\citealt{Forbes97}) has  blue, metal-poor GCs  formed during the collapse of the protogalactic cloud and associated to the galaxy halo. 
Red, metal-rich, GCs on the other hand might have been formed  during a  starburst that created the galaxy spheroid. Alternatively, \cite{shapiro10} suggest that red GCs might have formed in super star-forming clumps at $z\simeq2$ during the gas rich phase of galaxy evolution (see also  \cite{forbes15}). These clumps would have migrated toward the centre of the host galaxy, transporting GCs toward the proto-bulge. Some of the GCs would have been stripped from the migrating clumps and  become part of the newly born thick disk. Both those sub-populations would be old ($\ge 10$ Gyrs). In this simple scenario we expect to find a correlation between colour and kinematics of GCs: i.e. red GCs should share the same kinematics as spheroid or thick disk stars (\citealt{forbes12}) and blue GCs would trace the halo kinematics. Traditionally, this analysis has been performed dividing GCs according to their colour and then studying their kinematics (\citealt{forbes12,shapiro10}).
 In Section \ref{sec:analysis1} we reproduce such an analysis. In Section \ref{sec:model} we describe the galaxy model obtained from PNe and integrated stellar light.   In Section \ref{sec:analysis2}  we  present a new method that, instead, uses the GC kinematics to divide them into different sub-populations and then study their properties.  This method assumes that the galaxy halo is an extension of the galaxy bulge.  We call the bulge + halo components 'the galaxy spheroid', i.e. the spherically symmetric component dominated by random motion. This approximation could be relaxed when more data for the galaxy outskirts  become available.

\subsection{Radial distribution and kinematics of the red and  blue GCs}
\label{sec:analysis1}



\begin{figure}
\includegraphics[scale=0.45]{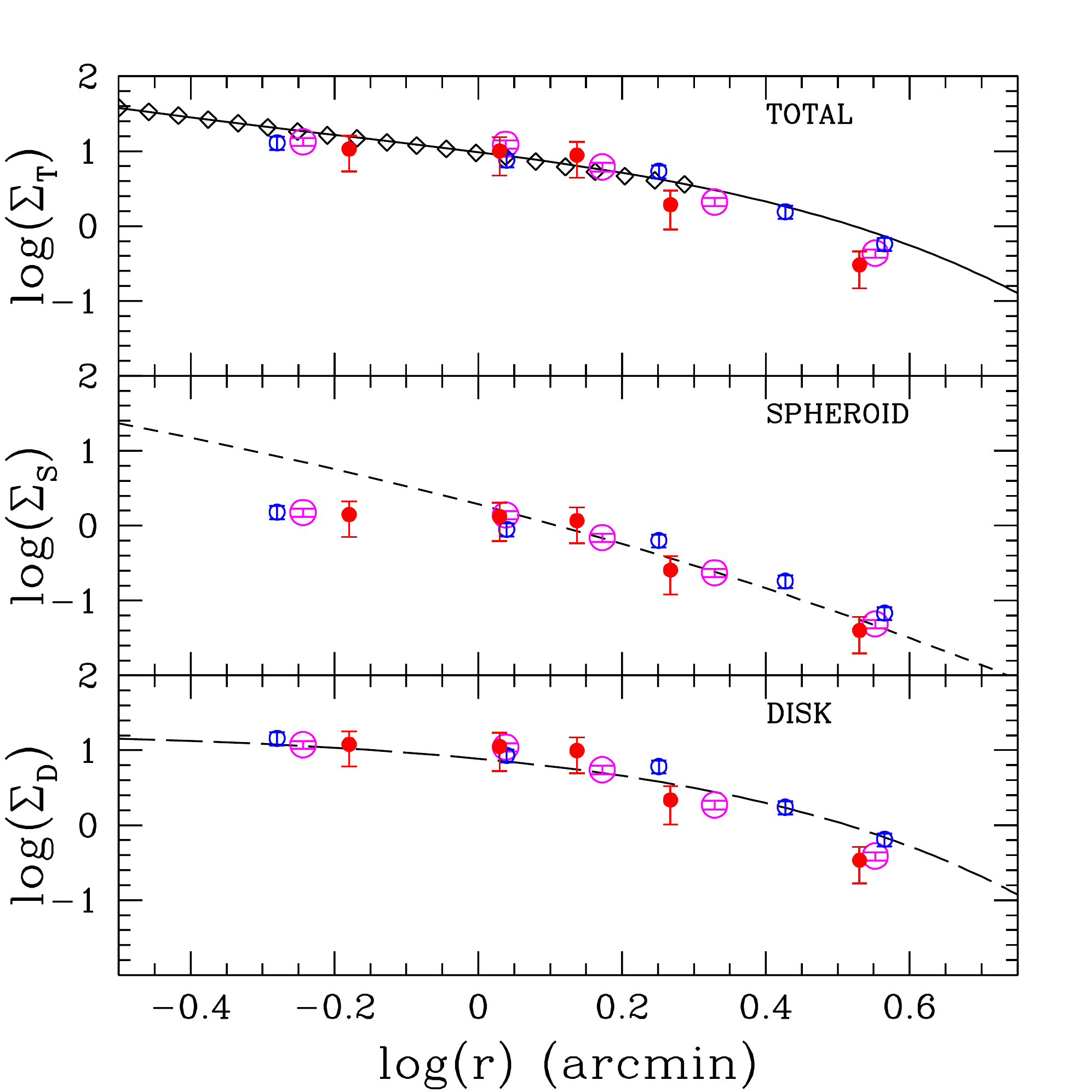}
\caption{Comparison of the galaxy surface brightness profile with the number density profile of the GCs. {\it Top panel}: The number density profile of {\it Photometric} GCs  in comparison with the galaxy R-band surface brightness from ellipse fitting (black diamonds)  and the galaxy light model from GALFIT (continuous black line). The number density profiles of the total GC population (large, magenta circles), red GCs (small, red, filled circles) and blue GCs (blue, open circles) are shown. {\it Middle panel}: Comparison is made with the spheroid model from GALFIT. {\it Bottom panel}: Comparison is made with the GALFIT disk model. In all panels, the data points have been shifted arbitrarily on the y-axis in order to match the galaxy light profile at $R=0.2$ arcmin. 
GCs follow the total light profile very well, the disk dominates the galaxy light. There is no visible difference between the two GC sub-populations. GCs might have been destroyed toward the galaxy centre or not been observed due to the bright galaxy background, therefore their distribution might be incomplete in the centre.}
\label{fig:Figure2}
\end{figure}

As a first analysis, we compare the one-dimensional radial distribution of GCs with the galaxy starlight: if GCs follow the underlying stellar population, then, to within a normalisation factor, the two might be expected to show the same profile. 
Figure \ref{fig:Figure2} shows the comparison of the total, blue and red {\it photometric GC} number densities (magenta, blue and red circles respectively) against the total, spheroid, and disk surface brightness (upper panel, middle panel, low panel, respectively), as modelled running GALFIT on an R-band image of the galaxy (\citealt{cortesi11}). The recovered parameters are: disk total absolute magnitude $7.02$ mag, disk scale length $59.08$ arcsec, disk axis ratio  $0.26$ and disk position angle $84.12$ degrees; for the spheroid we  estimated the total absolute magnitude  to be $6.9$ mag, effective radius $17.86$ arcsec, Sersic index $4$, axis ratio $0.75$, position angle  $75.59$. The spheroid over total light fraction is  $0.53$.
The GC number density has been obtained by  separating  the objects belonging to the main galaxy, NGC\,1023, from the objects associated to the companion galaxy, NGC\,1023A, using a light weighted probability (\citealt{Nor08}).  First, we bin the GCs of NGC\,1023 into elliptical annuli, going outward in radius, using the ellipticity obtained from a non parametric isophotal analysis of the R-band image of the galaxy. The total number of GCs in each bin is  then divided by the bin area. In Figure \ref{fig:Figure2}, the GC total, red and blue number densities have been shifted in the y-axis to match the modelled surface brightness profile. 
The total model, black continuous line in the upper panel of  Figure \ref{fig:Figure2}, describes well the light profile as obtained with an ellipse fitting (black diamonds) of the same R-band image. 
The total GC number density follows the total light profile, supporting the idea that they are associated with the galaxy. 
This galaxy light profile is dominated by the disk component and the GCs seem to follow the disk light profile. \cite{Kartha} found that the distribution of red and blue GCs in this galaxy is elongated along the major axis, resembling the disk isophotes, supporting this result. Nevertheless the spheroid profile would provide a good fit for the outer GCs, suggesting that red and blue GCs belong to both the galaxy components.  



\begin{figure*}
\includegraphics[width=1\textwidth]{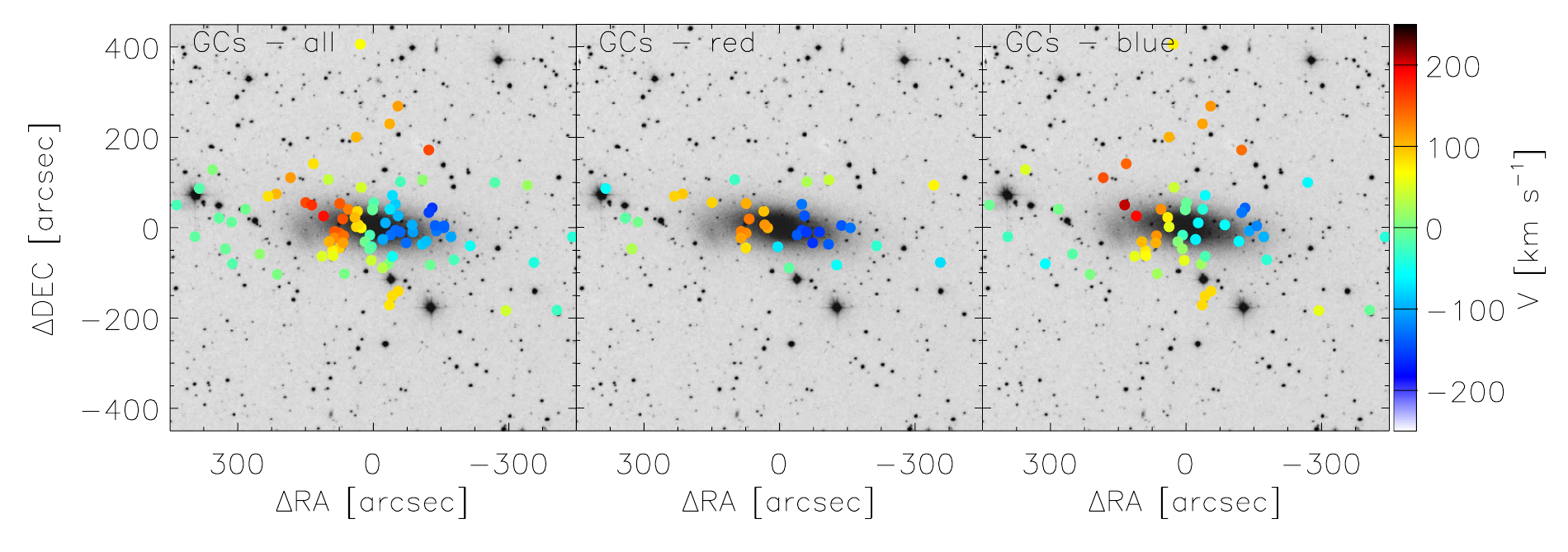}
\caption{2-D velocity maps of the GCs. Smoothed GC velocity field over-plotted on a DSS image of NGC~1023, \textit{left}, for the red sub-population \textit{centre} and the blue sub-population \textit{right}. Rotation $\simeq 180$ km/s is evident for both the red and blue sub-populations.}
\label{fig:Figure3}
\end{figure*}

A second clue comes from the study of the kinematics of the GC system as a whole,  and of the blue and red  sub-populations.
Figure \ref{fig:Figure3} shows the GC two-dimensional smoothed velocity fields, using an Adaptive Gaussian Kernel Smoothing (\citealt{c09}), for the total, red and blue sample  (left, middle, right  panel, respectively).
Rotation with a maximum velocity of  $180$ km/s, is clear in the total GC population as a whole, and for the red and the blue  sub-populations separately.  A similar result was found by  \cite{proctor08}, studying the GC system of the spiral galaxy NGC 2683.
The Milky Way red GCs also show large systemic rotation and were first associated with the old thick disk (\citealt{zinn85}) and then with the galactic bulge, due to the similarity of the net rotation of the innermost red GCs with that of the old field giant stars in the Galactic bulge (\citealt{zinn96}). 
\begin{figure*}
\includegraphics[width=1\textwidth]{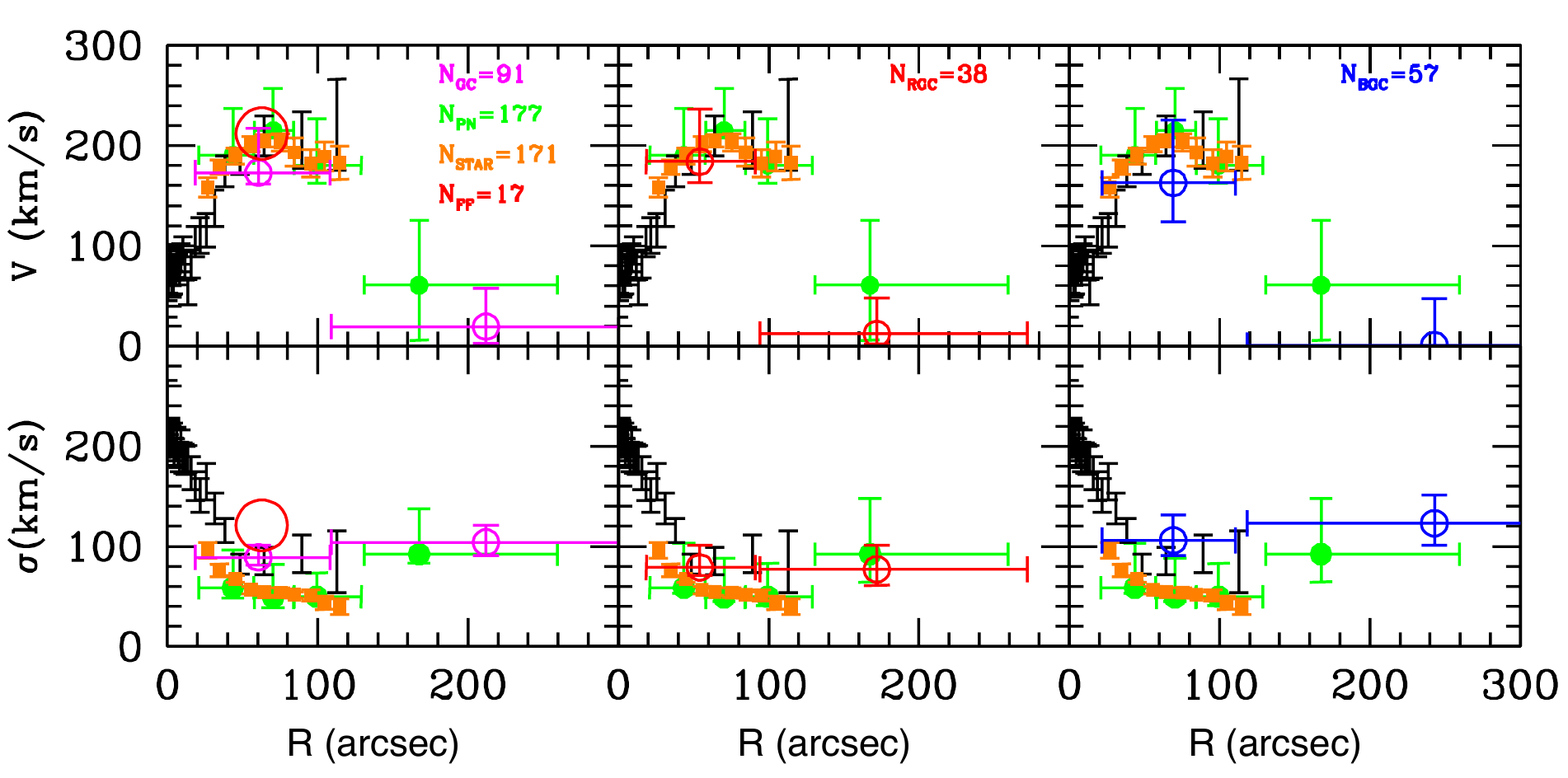}
\caption{Recovered kinematic profile for the GCs (magenta), PNe (green) and stellar data (orange). The top panels show the rotation velocity and the bottom panels the dispersion velocity. The radius has been obtained using the following formula $R^{2}=X^{2}*cos(I)+Y^{2}/cos(I)$, where $I$ is the inclination of the galaxy on the plane of the sky, estimated from the disk axis ratio fitted with GALFIT. Black open circles are absorption line data (\citealt{debattista02}).  The left panels include all the GCs and the FFs (red large open circle), the middle panels red GCs and the right panels blue GCs. All the kinematic tracers show the same kinematics profile. The kinematics of red and blue GCs are very similar within  errors.}
\label{fig:Figure4}
\end{figure*}
To compare the GC kinematics with those of the other tracers, we obtain the rotation velocity by fitting an inclined disk model using a maximum likelihood method (\citealt{cortesi11}) for  PNe and GCs, while the SKiMS kinematics are obtained with kinemetry (Foster et al., 2015, \textit{submitted}). 
The results are shown in Figure \ref{fig:Figure4}. All the tracers (even when dividing GCs into the blue and red sub-populations) have the unusual kinematic profile found by \cite{Nor08} for PNe, characterised by a decreasing rotation velocity and increasing random motions, in the outermost points (i.e. beyond 140 arcsec). 

The agreement found in  the spatial distribution and the kinematics of the three tracers suggests that we can use the stellar kinematics as recovered by integrated stellar light and PNe to model the GC velocity field.

\subsection{Kinematic model of NGC\,1023 from PNe and integrated stellar light}
\label{sec:model}

In Section \ref{sec:analysis1} we showed that, when recovering the kinematics of N1023  using  PNe, GCs and  integrated stellar light as tracers, the rotation velocity decreases at large radii, while the velocity dispersion increases. 
In a work based only on PNe data, \cite{cortesi11} explained this behaviour as a result of contamination of  the disk kinematics by PNe belonging to the spheroidal component. 
\cite{cortesi11} also show that assuming a spheroid + disk kinematic model, the unusual behaviour of the kinematics vanishes and the galaxy presents a spiral-like velocity profile, with a flat rotation curve, decreasing random motion in the disk and a pressure supported spheroidal component.
As a first test, we want to investigate whether, when a spheroid + disk model is applied, the SKiMS data also present  disk-like kinematics, as found for the PNe.
In the present section, we give a brief summary of the method used to obtain the spheroid and disk kinematics for PNe and integrated stellar light data, but we refer to \cite{cortesi11} for a more detailed explanation.

\begin{figure*}
\begin{tabular}{lr} 
\includegraphics[width=0.5\textwidth]{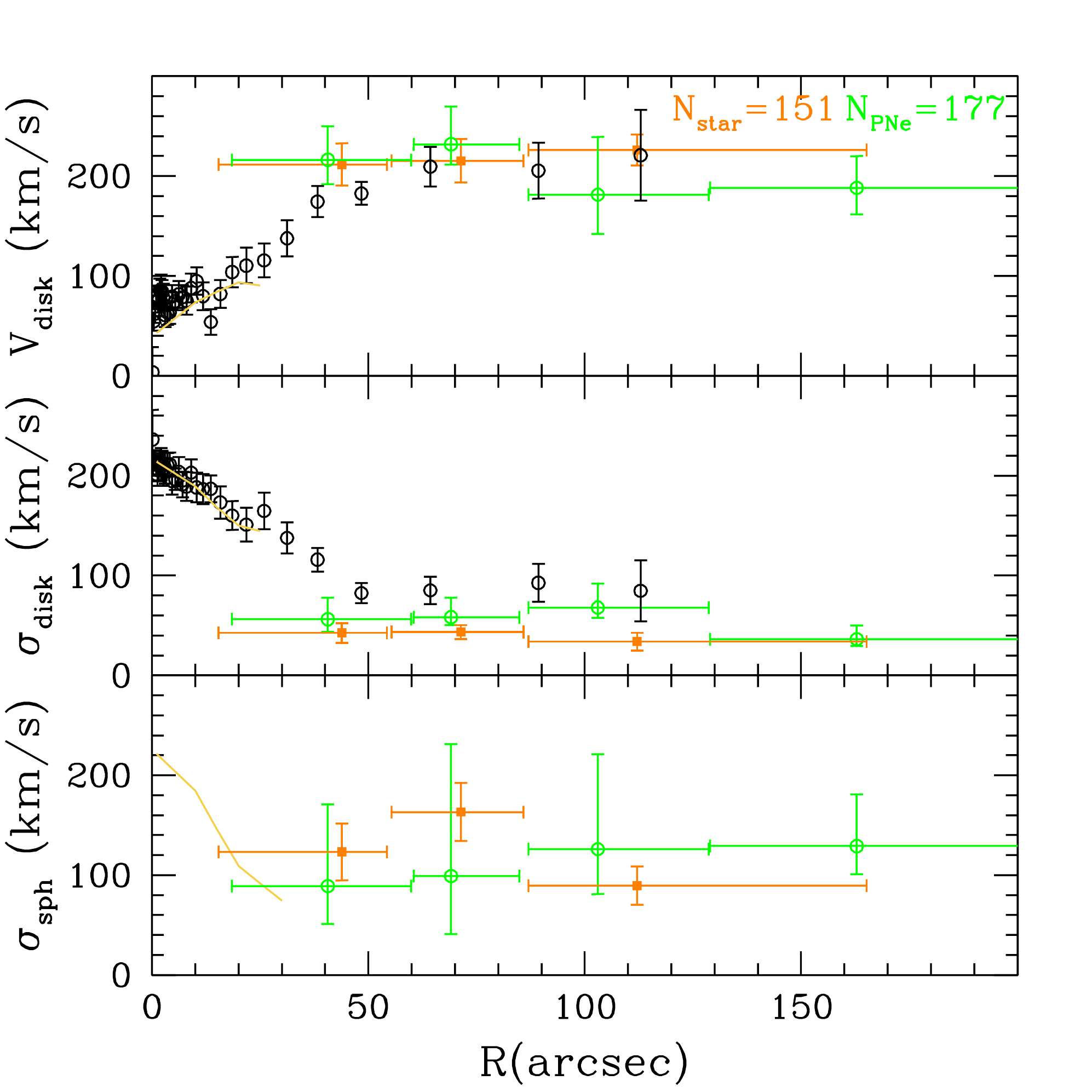} 
\includegraphics[width=0.5\textwidth]{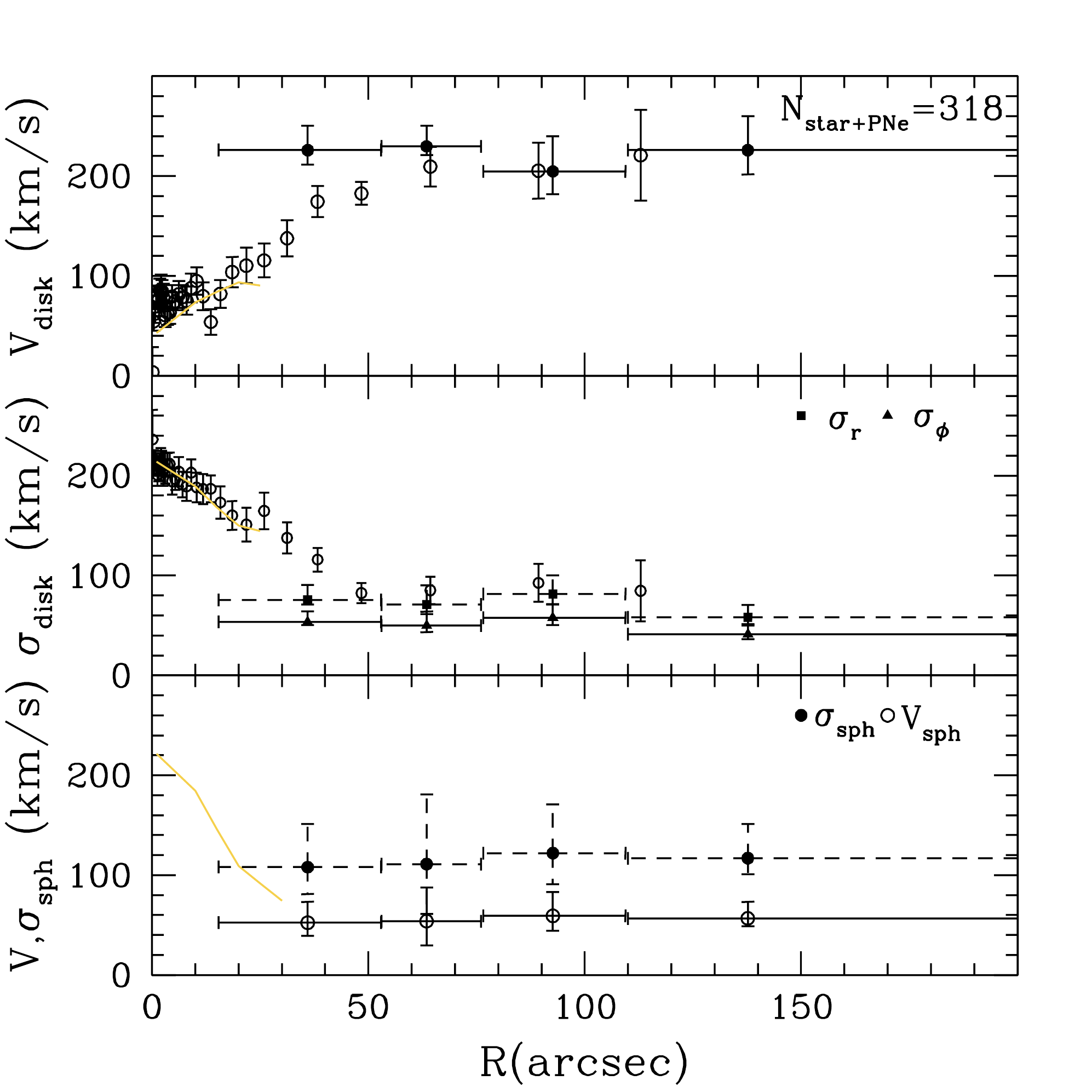} 
\end{tabular}
 \caption{NGC\,1023 disk and spheroid kinematics recovered with PNe and integrated stellar light. {\textit{Left panels}:} The best fit kinematic parameters for PNe (green dots) and stars (orange dots) are shown with 1$\sigma$ error bars. Error bars on the $x$ axis show the bin size.  Open, black circles correspond to absorption line data along the major axis (\citealt{debattista02}) and the yellow line is obtained by extracting SAURON data in cones along the major axis (top and middle panels) and minor axis (bottom panel). The top panel presents the disk rotation velocity, while the middle panel show the disk velocity dispersion. The bottom panel show the velocity dispersion in the spheroid. {\textit{Right}:} as in the {\textit{Left}} panel but  when the two data sets are unified in the same catalog.  Given the high number of tracers we can estimate the spheroid rotation, shown as black open dots with continuous error bars in the bottom panel.}
 \label{fig:Figure5}
\end{figure*}

 A galaxy R-band  image (\citealt{Nor08})  is decomposed into its spheroid and disk component using GALFIT (\citealt{peng02}). The PNe and SKiMS data are super-imposed on the image obtained by dividing the spheroid model by the total model, called f-map, and a probability $f_{i}$ (from 0 to 1) is assigned to each tracer; i.e. objects with $f_{i}=1$ would definitely be in the spheroid and object with $f_{i}=0$ would definitely be in the disk. We use a maximum likelihood method to estimate the best kinematics parameters (i.e. rotation velocity and velocity dispersion in the disk, dispersion velocity in the spheroid), assuming a velocity distribution function which is the sum of the spheroid distribution function (with zero mean rotation velocity) and the disk distribution function. The assumption that the mean spheroid rotation is zero will be relaxed below. Both distribution functions are assumed to be Gaussian and the contribution of each one of them to the total distribution is given by the parameter $f_{i}$. The fit is carried out in 4  radial bins for the PNe and  3 radial bins for the SKiMS data, with the same number of objects in each bin, always higher than 30 in order to assure a reliable fit. The fit is iterated several times, each time rejecting three-sigma outliers, i.e. objects with a likelihood lower than the {\it likelihood clipping probability threshold}, which is the value at which we cut the velocity distribution function (\citealt{cortesi11}). The fit rapidly converges rendering this process robust against a small amount of contamination. 
Since the SKiMS data are not point sources but slitlets, we have assumed that every slit is approximately described by a Gaussian velocity distribution function and we have extracted 10 random realisations  of  every slit, in order to have 10 different mock catalogs, and carried out the  analysis aforementioned on every catalog independently. In Figure \ref{fig:Figure5}, left panel,  the orange dots represent the median of the recovered parameters and the error is given by their standard deviation. The PNe kinematics are shown by  green dots. Error bars on the $y$-axis are    1 $\sigma$ error bars for the recovered parameters.  Error bars on the $x$-axis show the size of the bin where the fit was performed. The kinematics of stars and PNe is clearly similar. We can therefore unite the two data sets to  obtain a catalog of more than 300 tracers and recover the galaxy kinematics with more precision. Moreover, we estimate the rotation of the spheroid, see black circles with continuous error bars in the bottom, right panel of Figure \ref{fig:Figure5}. This estimate assumes that  $V_{sph}/\sigma_s \sim e / \sqrt{2}$, where $V_{sph}$ is the rotation velocity of the galaxy spheroid and $e$ the spheroid ellipticity (see Figure 4.14 in \citealt{binney87}). The GALFIT modelling shows that the spheroid in NGC\,1023 has an ellipticity of $\sim 0.25$, which translates into a predicted value of $V_s/\sigma_s \sim 0.5$.  We also recover two  components of the velocity dispersion $\sigma_{\phi}, \sigma_{r}$, see second row of Figure  \ref{fig:Figure5}, right panel (we assume $\sigma_{z} \simeq 0$). The average ratio between rotation velocity and velocity dispersion in the azimuthal direction, in the two outermost bins, is $\simeq 4.4 \pm 0.6$. This value is lower than the one expected for a spiral galaxy (\citealt{herrmann09}) suggesting that this S0 galaxy is not simply a faded spiral. Following \cite{bournaud05}, this value of $V / \sigma_{\phi}$ is consistent with this galaxy being involved in only very minor mergers in its past, see the Discussion section for more details. 


GCs give us, therefore, a unique opportunity for verifying the various possible formation histories of S0 galaxies, in particular  whether S0s are faded spirals or an independent class of galaxies. 
In fact, if the spiral galaxy has gone through a series of minor mergers that transformed it into an S0 galaxy, GCs may have been stripped from the accreted dwarf galaxies, which will pollute the chromo-dynamical relation found for spiral galaxies (\citealt{forbes01}). Nevertheless, we would expect to find the majority of  red GCs to be associated with the spheroid or thick disk and blue GCs with the galaxy halo, as found in spiral galaxies.  

\subsection{Chromo-dynamical analysis of GCs}
\label{sec:analysis2}

\begin{figure}
\begin{tabular}{c}
\includegraphics[width=0.45\textwidth]{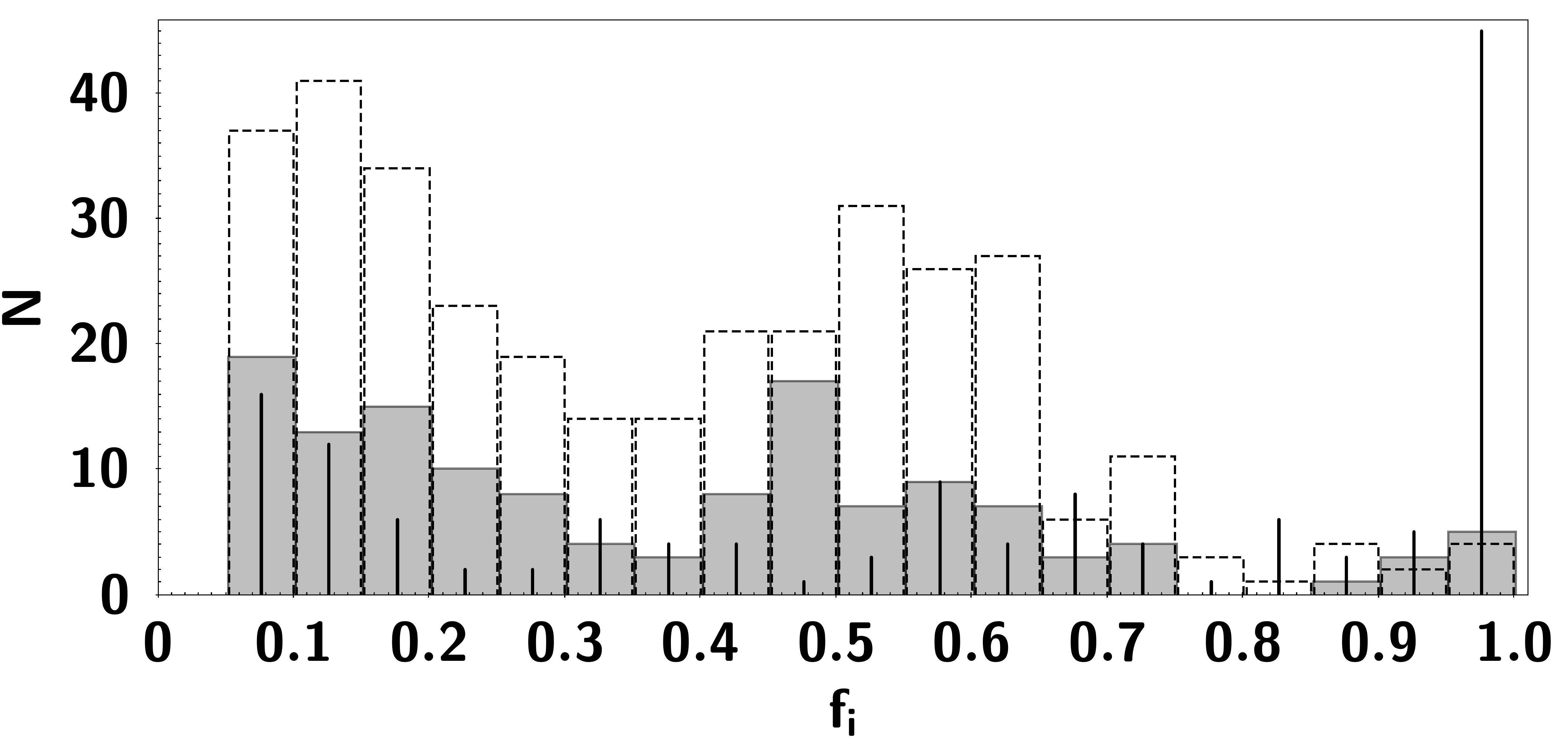}\\
\includegraphics[width=0.45\textwidth]{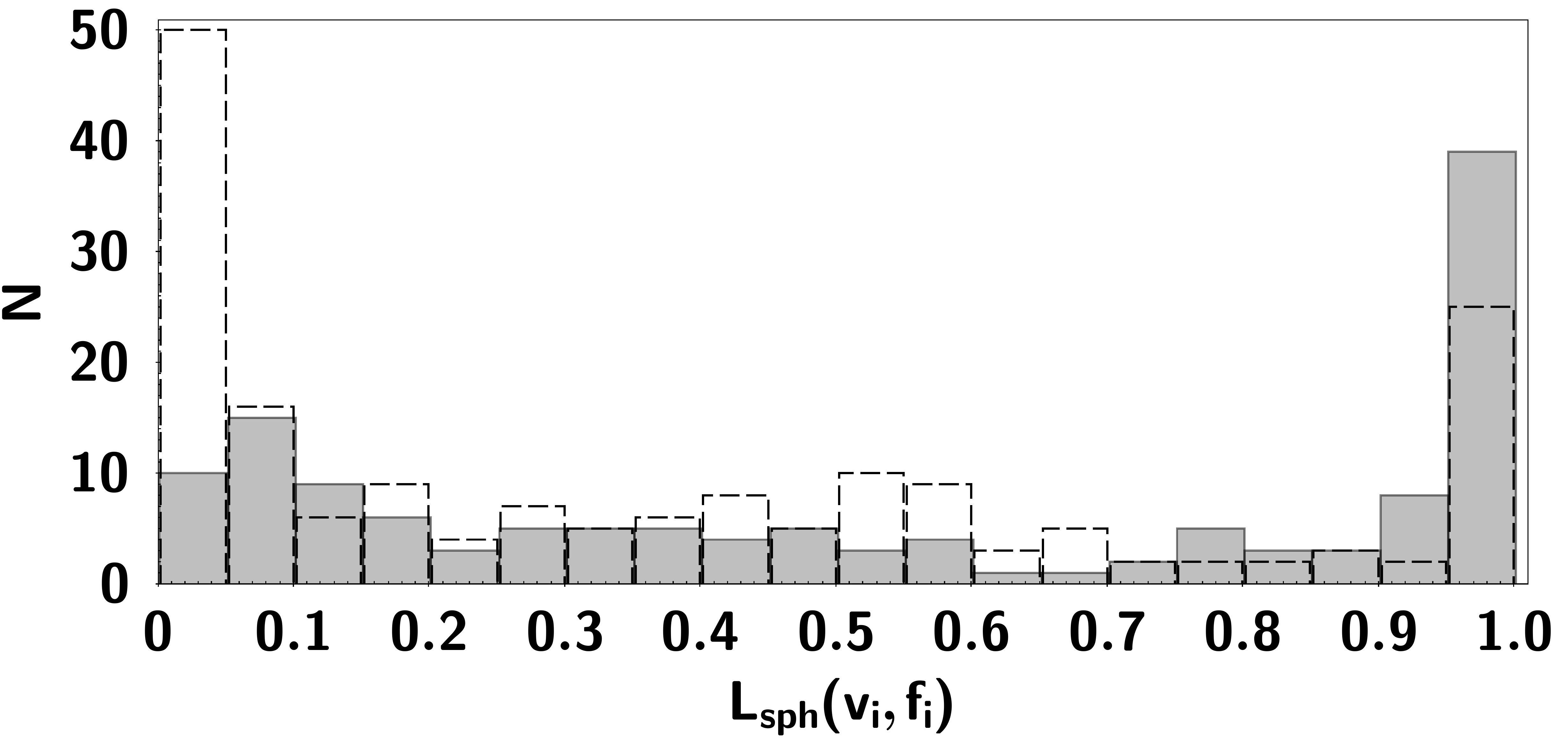}
\end{tabular}
\caption{Probability of belonging to the spheroidal component. {\it Upper panel}: histogram showing the probability of being in the spheroid  based on photometry $f_{i}$, as recovered over-plotting  on the f-map (see Section \ref{sec:model}) the GCs, filled histogram, the PNe, dashed histogram and a random distribution of object, with a 1/r density fall-off, black spikes. Objects with $f_{i}=1$ belong to the spheroid and objects with  $f_{i}=0$ belong to the disk. {\it Bottom panel}: histogram showing the probability of being in the spheroid based on photometry and kinematics, $\mathcal{L}_{sph}(v_{i},f_{i})$ for the GCs, filled area, and the PNe, dashed line.}
\label{fig:Figure6}
\end{figure}

In the previous section, using a unified catalog of  PNe and SKiMS datasets, we have recovered the  kinematics for the lenticular galaxy NGC\,1023, assuming a disk + spheroid model. Figure \ref{fig:Figure2} shows that in this galaxy the GCs follow the stellar light profile as modelled with GALFIT. 
It is therefore reasonable to use this model to assign a photometric-probability, $f_{i}$, to each GC of belonging to one or the other component.  As for the PNe and the integrated stellar light data, we obtained $f_{i}$  by over-plotting the GCs on the f-map (see Section \ref{sec:model}). In Figure \ref{fig:Figure6} upper panel, we show the histogram of the distribution of this parameter. It can be seen that the majority of the GCs appear to be associated with the disk component or have a probability around $0.5$ of being in the disk, as expected given the decomposition shown in Figure \ref{fig:Figure2} and the fact that the distribution of red and blue GCs in this galaxy is elongated along the major axis, resembling the disk isophotes \citep{Kartha}. In fact, when we get the probability of being in the spheroid for a random distribution of objects, whose density declines as $1/r$, we find that there is a higher number of objects associated with the spheroid, see the black spikes in Figure \ref{fig:Figure6} upper panel.

Given the best estimators  $V, \sigma_{r}, \sigma_{\phi},  \sigma_{s}$,  within each radial bin, obtained via a maximum likelihood fitting of PNe and SKiMS data (see Section \ref{sec:model}) and the parameter $f_{i}$  described in the previous paragraph, we can write the likelihood of each GC to belonging to the system as: 

  \begin{eqnarray}
\lefteqn{\mathcal{L}(v_{i},f_{i};V,\sigma_{r},\sigma_{\phi},\sigma_{s}) \propto  
      {f_{i} \over \sigma_s} \exp \left[ - \frac{v_{i}^{2}}{2 \sigma_{s}^{2}} \right]}  \nonumber \\ 
 & & + {{1-f_{i}} \over \sigma_{los}}\exp \left[ - \frac{(v_{i} - {V_{los}})^{2}}{2\sigma_{los}^{2}} \right], 
\label{eq:likelihood}
\end{eqnarray}
where 
\begin{equation}
V_{los}=V_{sys}+V \sin(i) \cos(\phi), 
\end{equation}
is the projection of the galaxy's mean rotation velocity $V$,
\begin{equation}
\sigma^{2}_{los}=\sigma^{2}_{r} \sin^2 i \sin^2\phi +\sigma^{2}_{\phi} \sin^2i \cos^2\phi + \sigma^{2}_{z} \cos^2i,
\end{equation}
is the line-of-sight velocity dispersion in cylindrical polar coordinates  ($R,\phi,z$), assuming that  $\sigma_z$  is negligible  in its contribution to $\sigma_{los}$, because of the near edge-on viewing angle,  and $\sigma_{s}$ is the dispersion in the spheroid. 
Equation \ref{eq:likelihood} gives us the probability for each GC of belonging to the NGC\,1023 system. If the recovered total likelihood is lower than the  {\it likelihood clipping probability threshold},  the GC is labeled as unlikely to belong to the system at $2.1\sigma$ level of confidence (see Section \ref{sec:model} and \citealt{cortesi11} for details). This threshold has been chosen for consistency, since it is the same one used to fit the PNe and stellar kinematics. In \cite{cortesi11}, different level of confidences have  been tested on a model galaxy obtained from a self-consistent N-body simulation and  $2.1\sigma$ was selected for providing a robust fit.
The likelihood of being in the spheroid, $\mathcal{L}_{sph}(v_{i},f_{i})$  is proportional to the first part of Equation \ref{eq:likelihood}   normalised by the total likelihood.  The likelihood of being in the disk, $\mathcal{L}_{disk}(v_{i},f_{i})$, is proportional to the second part of Equation \ref{eq:likelihood}  normalised for the total likelihood. For each GC given its position and velocity and its photometric probability of being in the spheroid, we can estimate its likelihood of belonging to the system  and of being in one of the two components.
 The distribution of the probability of being in the spheroid, $\mathcal{L}_{sph}(v_{i},f_{i})$, obtained combining photometry and kinematics (lower panel of Figure \ref{fig:Figure6}) is clearly different from that recovered only from photometry.  In particular, the number of objects for which the probability of being in the disk is equal to the probability of being in the spheroid has decreased. In fact, assigning objects to different components only on the basis of the photometric decomposition is inadequate, due to projection effects. Combining photometric information with kinematics  improves our capacity to discriminate between spheroid and disk objects, since it adds  constraint where the photometric probability of belonging to the disk is equal to the probability of being on the spheroid, i.e. the probability is 50\% of being in the disk and 50\% of being in the spheroid. Moreover, objects that lie along the major axis, and would be therefore assigned to the disk on the basis of photometry, but present no rotation or are conter-rotating are rejected. Note that the recovered likelihood of being in the spheroid for GCs is very similar to the likelihood of being in the spheroid obtained for the PNe, i.e. the majority of the objects are associated with the spheroid or the disk, see Figure \ref{fig:Figure6}, since combining photometry with kinematics breaks down the degeneracy due to projection effects.  The number of PNe associated to the disk is higher than the number of PNe associated with the spheroid; the contrary is true for the GCs.
The method is in general applicable to every discrete kinematic tracer and other components can be added to the model as a thin disk or a halo, if the necessary data are available. 
The recovered likelihood is listed in the GC catalog published with this paper.

\section{Results}
Having computed the likelihood for every GC of belonging to the model of the galaxy obtained combining photometric information from an R-band galaxy image and PNe and SKiMs kinematics, in this section we study the properties of the GC population and sub-populations. We have  defined four different sub-populations: red and blue GCs, on one hand, and disk and spheroid GCs, on the other. It should be noted that in this preliminary study the spheroid may incorporate bulge and halo sub-populations with  different properties. In Section \ref{sec:kinecolor} we investigate the relation between these four sub-populations and the galaxy as a whole. In Section \ref{sec:rej} we study all the kinematic tracers unlikely to belong to NGC\,1023 system and their connection with the companion galaxy, NGC\,1023A.
 
\subsection{Kinematic - colour relation}
\label{sec:kinecolor}

\begin{figure*}
 \includegraphics[width=0.9\textwidth]{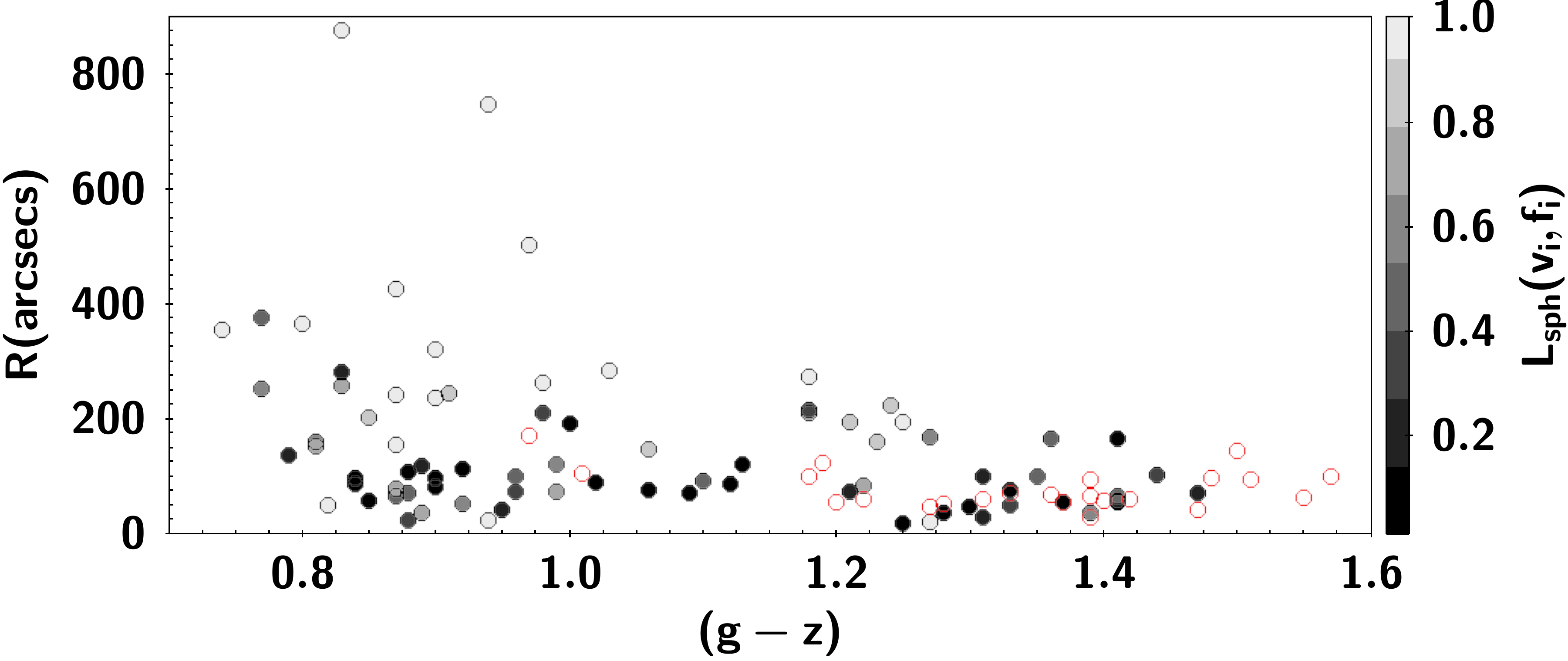}
 \caption{GC colour versus distance from the galaxy centre. The GCs are colour coded according to their probability of   belonging to the spheroid. Large, red, open circles mark the position of the FFs in the same plane.  The reddest GCs are likely to be associated to the galaxy disk and share the same loci as FFs in this plot. The majority of blue GCs are likely to be associated with the spheroid, but  some have a high probability of being in the disk.}
\label{fig:Figure7}
\end{figure*}

To obtain a clearer picture of the correlation between GC colour and kinematics, we have to take into consideration the shape of the bi-modal colour distribution. In fact, the red and blue sub-populations overlap at intermediate  colours, and our artificial colour cut at $(g-z)=1.1$ is an approximation. We fitted the two sub-populations with a Gaussian function and  estimated that the area in which the two sub-populations significantly overlap is  $1.0< (g-z) <1.1$ and we exclude GCs with  these colours from our analysis. As a first step, we calculate the total number of disk and spheroid GCs, simply summing up the probability for each one of them of being in the disk or in the spheroid. We find that out of the 32 red GCs, 19 are associated with the disk and 13 with the spheroid. While  18.6 blue GCs are associated with the disk and 25.4 with the spheroid. Finally, 15 GCs are found to have a probability lower than 2.1 $\sigma$ of belonging to the system. The majority of the rejected GCs have blue colours.
 Looking at the trend  of $\mathcal{L}_{sph}(v_{i},f_{i})$ vs the (g-z) colour of the GCs, we realise that an extra parameter is needed to make this plot meaningful. Figure \ref{fig:Figure7} shows the distribution of the GCs (associated with the system) in a colour vs galactocentric distance plot, colour coded according to the probability of being in the spheroid $\mathcal{L}_{sph}(v_{i},f_{i})$.    
It is clear that the reddest GCs are concentrated in the centre of the galaxy and that they are associated with the galaxy disk. This fact is of particular interest since in this galaxy a population of FFs has been found to have rotational velocity similar to the stellar one (\citealt{LB00}). FFs are also shown in Figure \ref{fig:Figure7} as red open circles. They share the same  position as the disk red GCs. For red GCs it seems that the reddest the GC the higher is its probability of belonging to the disk. For blue GCs, instead, this probability seems to correlate with its distance from the galaxy centre, with GC at  low radii having a higher probability of belonging to the disk.

\subsection{Rejections and streams}
\label{sec:rej}

\begin{figure*}
 \includegraphics[width=0.45\textwidth]{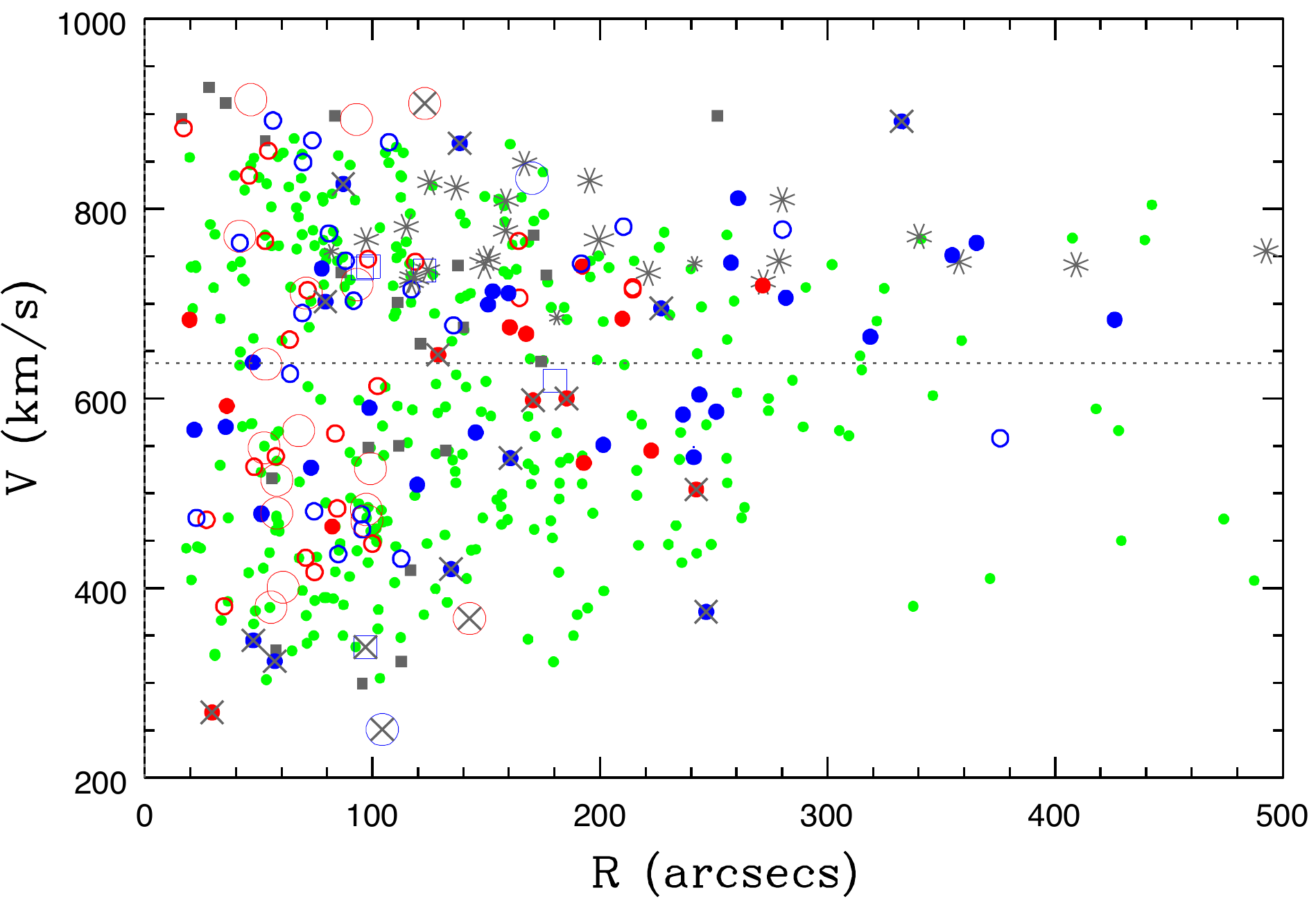}\includegraphics[width=0.45\textwidth]{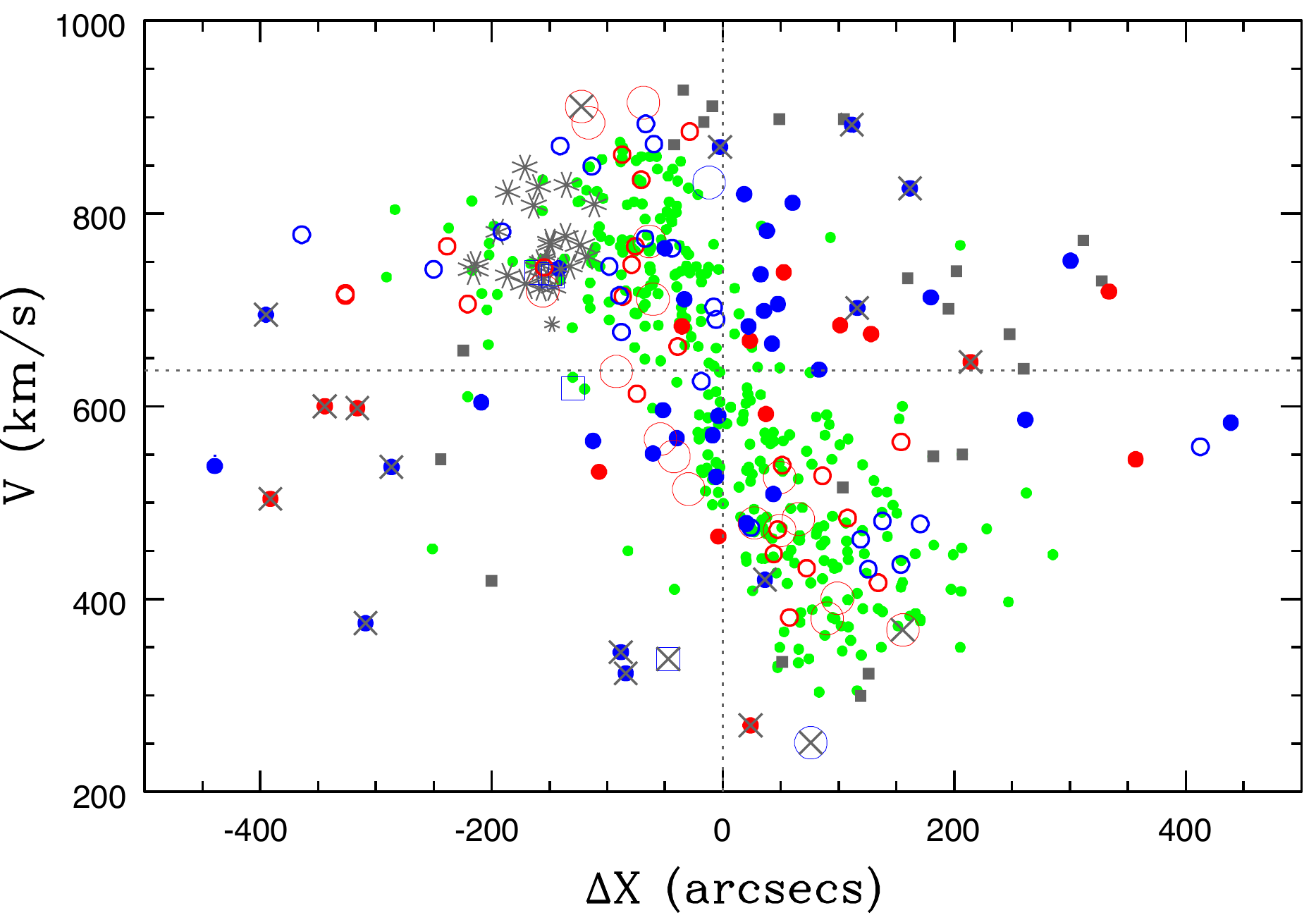}
 \caption{ {\it Left:} GCs 1-D phase space diagram. Rejected objects are shown as grey symbols (GCs; crosses, PNe and stars; squares and companion galaxy objects; asterisks). Objects included in our model are shown as coloured symbols. For GCs, small open circles identify objects belonging to the disk, while small filled circles identify objects belonging to the spheroid.  UCDs  are shown as blue squares, FFs as large circles and PNe and integrated stellar light individual data points, as green dots.  The dashed horizontal line shows the galaxy systemic velocity. The higher the distance from the galaxy systemic velocity, lower is the infall time, see text.  { \it Right:} Position along the major axis vs velocity. Symbols are like in the left panel. The vertical line shows the centre of major axis. Rejected objects form a coherent structure connected to the companion galaxy.}
\label{fig:Figure8}
\end{figure*}

 The method used in this paper to recover the galaxy kinematics allows us to identify outliers, i.e. objects that are not in equilibrium with  NGC\,1023  disk + spheroid potential (see \citealt{cortesi11} for details). 
Some of the rejected GCs fall in an area of the galaxy where there are no data available to recover the kinematics. Therefore we have extrapolated  the galaxy kinematics in these outer regions from the outermost bin, for which we have PNe and stellar kinematics. The outer parts of this   galaxy are generally dominated by the spheroid component that we assume to  represent the bulge + halo components of the galaxy. If this assumption is not correct these outliers could be halo GCs. On the other hand, they could be  objects that are in  the process of being accreted to the system. 

\cite{Rocha}, in a study of the kinematics of the Milky Way satellites, shows that it is possible to quantify the time since they have started falling onto the MW using a  1-D phase space diagram. In such a diagram  the infall time of an object is proportional to its  distance from the galaxy systemic velocity.  Figure \ref{fig:Figure8}, left panel,  shows the 1-D phase space diagram for NGC\,1023. Rejected objects are shown as grey symbols (GCs; crosses, PNe and stars; squares and companion galaxy objects; asterisks). Objects included in our model are shown as coloured symbols. The PNe and integrated stellar light individual data points are shown as tiny green squares, GCs as small circles and FFs as large circles. The galaxy systemic velocity is shown as a dashed line. PNe, integrated stellar light data points, and   GCs form a half-diamond structured centred around the galaxy  systemic velocity. The companion galaxy PNe and GCs and some of the rejected objects lie at the borders of this half-diamond, as if they are just starting in-falling. Figure \ref{fig:Figure8}, right  panel, shows the velocity vs distance along the major axis for all the objects, following the same symbols' convention as in the left panel. 
The PNe and stars follow a radial gradient in velocity, typical of  rotating system. 
 GCs and FFs follow the same distribution as PNe. In particular, the GCs belonging to the disk (open small circle)   lie in the external part of the distribution, i.e. they have a high velocity respect to the galaxy systemic velocity. The majority of the GCs associated with the galaxy spheroid (filled small circle), instead, are located near the galaxy centre or along the minor axes.
The rejected objects seem to form a coherent structure in the phase-space, connected with the companion galaxy, suggesting  they may have been stripped from the companion while orbiting the main galaxy. Three of the UCDs are in the same locus as the companion galaxy and  one belongs to the stream, suggesting a possible origin of the UCD connected with the interaction of the two galaxies  (\citealt{duc97}). This result needs confirmation with more data. 

Figure \ref{fig:Figure8} support a scenario where the rejected GCs are part of a stream connecting NGC\,1023A with the main galaxy. This possibility can be fully confirmed when data to properly describe the galaxy halo will be available.


\section{Discussion}
In this paper we present a new catalog of spectroscopically confirmed GCs associated with  the lenticular galaxy NGC\,1023 and its companion galaxy NGC\,1023A. We find that in this galaxy the GCs follow the total galaxy light profile and are therefore good tracers of the overall stellar population. Their kinematics are also consistent with the stellar kinematics. We divide the GCs according to their colour, at $(g-z)=1.1$, and find that the light profile and kinematics of both the red and blue sub-populations also resemble the stellar light profile and kinematics. 

However, for both red and blue GCs, it is not clear from the comparison of the number density profiles with the decomposed disk and spheroid profiles whether the GCs better match either the disk or spheroid light profiles. 
We, therefore, develop a new method to associate GCs to the galaxy disk and spheroid on the basis of their position and velocities, using PNe, integrated stellar light and an R-band image of the galaxy, to model the galaxy kinematics. 
We find that  the majority of red GCs (19) have a high likelihood of belonging to the galaxy disk. The other (13) red GCs  belong to the spheroid.
 Interestingly NGC\,1023 has a sub-population of red FFs (\citealt{LB00}) whose kinematics  are indistinguishable from the stellar kinematics of the disk (\citealt{cs13}) and whose azimuthal distribution is elongated along the galaxy major axis (\citealt{forbes14}). 
 The majority of blue GCs are spheroid-like objects (25.4), but, surprisingly $18.6$ are found to be disk-like objects.

The ratio between rotation velocity and velocity dispersion in the azimuthal direction in NGC\,1023 is $\simeq 4.4 \pm 0.6$. This value is lower than the one expected for  spiral galaxies (\citealt{herrmann09}) suggesting that this S0 galaxy is not simply a faded spiral. Following \cite{bournaud05}, this value of $V / \sigma_{\phi}$ is consistent with this galaxy having  been involved in only very minor mergers in its past.
An alternative explanation for our finding that $V/\sigma$ is rather low in the disk of N1023 is provided by the possibility of forming the galaxy disk at high redshift through the  merger of  super star-forming clumps during the gas rich phase of galaxy evolution.
\cite{shapiro10}, in fact, suggested that red GCs are born within `turbulent and clumpy' disks at $z \simeq 2$, i.e. disks forming in a violent way.  At $z \simeq 0$, these disks (when they survive) may look much hotter than the thin disks in spiral galaxies.
Clumpy disk formation, moreover, would leave imprints in the halo, as suggested by simulations from \cite{inoue13}. Halo objects could interact gravitationally with the giant clumps and acquire  rotation in a dynamical friction timescale of 0.5\,Gyr. The influence of the clumps would be limited to a region around the disk. This formation scenario  explain the finding that red and blue GCs in the inner part of the galaxy have a disk-like kinematics.
Nevertheless,  the galaxy is going through an interaction (and maybe a  minor merger) at the present epoch with  a companion galaxy, NGC\,1023A,  visible inside the galaxy disk optical isophotes. This little companion may have been stripped of its gas, producing  an  extended and irregular HI distribution, displaced from the optical centre (\citealt{sancisi85}).  We find that some of the GCs that are  unlikely to be in equilibrium with the potential of NGC\,1023 are  part of a stream, probably connected to the companion galaxy. 
Blue FFs, moreover, are associated with the companion galaxy and the HI clumps (\citealt{forbes14}). 
Some of the GCs unlikely to belong to the model of the galaxy could be halo GCs that are not well represented by our simple model and are, therefore, rejected during the fitting.  

NGC\,1023  seems to have a population of GCs associated with its disk. In contrast,  in  the S0 galaxy NGC\,2768 red GCs share the same kinematics and density profile as spheroid stars (\citealt{forbes12}), as found for the MW and the Sombrero galaxy. This framework  provides an interesting possibility for testing how a given galaxy disk formed: by seeing if there are associated red GCs.

As more GC data in spiral galaxies become available, together with the ages of S0 galaxy disk and spheroid, and PNe kinematics of galaxy halos are obtained, the formation scenario for  S0 galaxies will become clearer.
For the moment, we find that GCs are powerful probes of galaxy evolution and that the red GCs in the two lenticular galaxies NGC\,1023 and NGC\,2768 are associated with different galaxy components, suggesting that S0 galaxies form through different pathways (or that their progenitors formed through different pathways).
\section{Summary and Conclusions}
In this paper we have published a new catalog of GCs  and introduced a new method to associate GCs with different galactic components, given a kinematic model. This method also  allows us to identify streams and can be applied to any discrete kinematic tracers. In particular we have used PNe, stellar light and an R-band image to constrain the kinematic model of NGC\,1023, decomposing it into its disk and spheroid components. Following that we have studied its GC population. The main findings are:
\begin{itemize}
\item The GCs follow the stellar density profile and kinematics.
\item Both red and blue GC sub-populations show  rotation.
\item When decomposing  the galaxy into its spheroid and disk component, we find that the majority of the red GCs show disk-like kinematics, the others are consistent with being in the spheroid. Most of the blue GCs have a high likelihood to belong to the spheroid, but some of them show disk-like kinematics. 
\item The GCs that are unlikely to belong to the galaxy model appear to be associated with a  stellar stream, connected to the companion galaxy NGC~1023A, previously identified by \cite{cortesi11} using PNe. We note that   they could also be halo GCs that are not well represented by our simple model and are, therefore, rejected during the fitting.
\end{itemize}
All these findings are consistent with this galaxy having been involved in at least one minor mergers, while its disk was created at high redshift from super-star-forming clumps. Moreover, in the only other galaxy for which similar data are available, NGC\,2768,  the red GCs share the same kinematics and density profile as spheroid stars. This suggests that S0 galaxies are an heterogeneous class and that GCs are  powerful probes of galaxy evolution.  The newly developed method will be applied on a sample of lenticular galaxies to place stronger  constraints in the galaxy-globular clusters assembly history.  

\section*{Acknowledgments}
The authors wish to thank the anonymous referee for useful comments.
The authors would like to thank Dr. D. Kruijssen and Dr. Y. Jaffe for interesting discussions and suggestions. This work was supported by NSF grant AST-1211995.  AC would like to thank  FAPESP for the  fellowship 2013/04582-4.  ACS acknowledge receipt of a Royal Society travel grant and the hospitality at UC Santa CruzACS and funding from a CNPq, BJT-A fellowship (400857/2014-6). DAF thanks the ARC for financial support via DP130100388. RP acknowledges the support of CNPq through grant 157761/2014-2.

\label{lastpage}


\begin{thebibliography}{99}
\providecommand{\gca}{Geochim.~Cosmochim.~Acta}
\providecommand{\aaps}{\mbox{AAPS}}
\providecommand{\pasj}{\mbox{PASJ}}
\providecommand{\pasp}{\mbox{PASP}}
\providecommand{\apj}{\mbox{ApJ}}
\providecommand{\apjs}{\mbox{ApJS}}
\providecommand{\apjl}{\mbox{ApJL}}
\providecommand{\araa}{\mbox{ARA\&A}}
\providecommand{\aj}{\mbox{AJ}}
\providecommand{\mnras}{\mbox{MNRAS}}
\providecommand{\aap}{\mbox{A\&A}}
\providecommand{\nat}{\mbox{Nature}}
\providecommand{\aplett}{\mbox{Astrophysical Letters}}
\providecommand{\prc}{Physical Review C}
\bibitem[Arag{\'o}n-Salamanca et 
al.(2006)]{salamanca06} Arag{\'o}n-Salamanca, A., Bedregal, A.~G., \& Merrifield, M.~R.\ 2006, \aap, 458, 101 

\bibitem[Arag{\'o}n-Salamanca(2008)]{Salamanca} 
Arag{\'o}n-Salamanca, A.\ 2008, IAU Symposium, 245, 285 

\bibitem[Alamo-Mart{\'{\i}}nez et al.(2013)]{alamo-mart13} 
Alamo-Mart{\'{\i}}nez, K.~A., Blakeslee, J.~P., Jee, M.~J., et al.\ 2013, 
\apj, 775, 20 

\bibitem[Arnold et al.(2014)]{arnold13} Arnold, J.~A., 
Romanowsky, A.~J., Brodie, J.~P., et al.\ 2014, \apj, 791, 80 




\bibitem[Binney \& Tremaine, 1987]{binney87} Binney, J., \& Tremaine,
  S.\ 1987, Princeton, NJ, Princeton University Press, 1987, 747 
  
\bibitem[Borlaff et al.(2014)]{Borlaff2014} Borlaff, A., 
Eliche-Moral, M.~C., Rodr{\'{\i}}guez-P{\'e}rez, C., et al.\ 2014, 
arXiv:1407.5097 

\bibitem[Bournaud et al., 2005]{bournaud05} Bournaud, F., Jog, C.~J., \&
  Combes, F.\ 2005, \aap, 437, 69


\bibitem[Brodie 
\& Strader(2006)]{bs06} Brodie, J.~P., \& Strader, J.\ 2006, \araa, 44, 193 

\bibitem[Brodie et al.(2012)]{brodie12} Brodie, J.~P., Usher, 
C., Conroy, C., et al.\ 2012, \apjl, 759, LL33 

\bibitem[Brodie et al. (2014)]{Brodie14}
{Brodie} J.~P.,  {Romanowsky} A.~J.,  {Strader} J.,  {Forbes} D.~A.,  {Foster}
  C.,  {Jennings} Z.~G.,  {Pastorello} N.,  {Pota} V.,  {Usher} C.,  {Blom} C.,
   {Kader} J.,  {Roediger} J.~C.,  {Spitler} L.~R.,  {Villaume} A.,  {Arnold}
  J.~A.,  {Kartha} S.~S.,    {Woodley} K.~A.,  2014, \apj, 796, 52
  

  
\bibitem[Byrd \& Valtonen(1990)]{Byrd} Byrd, G., \& Valtonen, M.\ 1990, \apj, 350, 89 

\bibitem[Cantiello et 
al.(2014)]{Cantiello} Cantiello, M., Blakeslee, J.~P., Raimondo, G., et al.\ 2014, \aap, 564, L3 



\bibitem[Chies-Santos et 
al.(2011)]{CS11b} Chies-Santos, A.~L., Larsen, S.~S., Kuntschner, H., et al.\ 2011, \aap, 525, A20 



\bibitem[Chies-Santos et 
al.(2012)]{cs12a} Chies-Santos, A.~L., Larsen, S.~S., Cantiello, M., et al.\ 2012, \aap, 539, A54 

\bibitem[Chies-Santos et 
al.(2013)]{cs13} Chies-Santos, A.~L., Cortesi, A., Fantin, D.~S.~M., et al.\ 2013, \aap, 559, AA67 



\bibitem[Coccato et al.(2009)]{c09} Coccato, L., Gerhard, 
O., Arnaboldi, M., et al.\ 2009, \mnras, 394, 1249 


\bibitem[Cortesi et al.(2011)]{cortesi11} Cortesi, A., 
Merrifield, M.~R., Arnaboldi, M., et al.\ 2011, \mnras, 414, 642 

\bibitem[Cortesi et al.(2013a)]{cortesi13a} Cortesi, A., Arnaboldi, M., Coccato, L., et al.\ 2013, \aap,549, AA115 

\bibitem[Cortesi et al.(2013b)]{cortesi13b} Cortesi, A., 
Merrifield, M.~R., Coccato, L., et al.\ 2013, \mnras, 432, 1010

\bibitem[Debattista et al.(2002)]{debattista02} Debattista, V.~P., 
Corsini, E.~M., \& Aguerri, J.~A.~L.\ 2002, \mnras, 332, 65

\bibitem[Desai et al.(2007)]{desai07} Desai, V., Dalcanton, 
J.~J., Arag{\'o}n-Salamanca, A., et al.\ 2007, \apj, 660, 1151 

\bibitem[Douglas et al.(2002)]{douglas02} Douglas, N.~G., 
Arnaboldi, M., Freeman, K.~C., et al.\ 2002, \pasp, 114, 1234 

\bibitem[Dressler et al.(1997)]{dressler97} Dressler, A., Oemler, 
A., Jr., Couch, W.~J., et al.\ 1997, \apj, 490, 577 

\bibitem[Dressler \& Sandage(1983)]{dressler83} Dressler, A., \& Sandage, A.\ 1983, \apj, 265, 664 

\bibitem[Duc 
\& Mirabel(1994)]{duc97} Duc, P.-A., \& Mirabel, I.~F.\ 1994, \aap, 289, 83 

\bibitem[Forbes et al.(1997)]{Forbes97} Forbes, D.~A., Brodie, 
J.~P., \& Grillmair, C.~J.\ 1997, \aj, 113, 1652 

\bibitem[Forbes et al.(2001)]{forbes01} Forbes, D.~A., Brodie, 
J.~P., \& Larsen, S.~S.\ 2001, \apjl, 556, L83 

\bibitem[Forbes et al.(2012)]{forbes12} Forbes, D.~A., Cortesi, 
A., Pota, V., et al.\ 2012, \mnras, 426, 975


\bibitem[Forbes et al.(2014)]{forbes14} Forbes, D.~A., Almeida, 
A., Spitler, L.~R., \& Pota, V.\ 2014, \mnras, 442, 1049 

\bibitem[Forbes et al.(2015)]{forbes15} Forbes, D.~A., 
Pastorello, N., Romanowsky, A.~J., et al.\ 2015, arXiv:1506.06820 


\bibitem[Foster et al.(2011)]{foster11} Foster, C., Spitler, 
L.~R., Romanowsky, A.~J., et al.\ 2011, \mnras, 415, 3393 

\bibitem[Foster et al.(2013)]{foster13} Foster, C., Arnold, 
J.~A., Forbes, D.~A., et al.\ 2013, \mnras, 435, 3587 



\bibitem[Gunn \& Gott(1972)]{Gunn} Gunn, J.~E., \& Gott, J.~R., III 1972, \apj, 176, 1 



\bibitem[Harris(2010)]{harris10} Harris, W.~E.\ 2010, Royal 
Society of London Philosophical Transactions Series A, 368, 889 

\bibitem[Herrmann 
\& Ciardullo(2009)]{herrmann09} Herrmann, K.~A., \& Ciardullo, R.\ 2009, \apj, 705, 1686 

\bibitem[Inoue(2013)]{inoue13} Inoue, S.\ 2013, \aap, 550, A11 




\bibitem[Kartha et al.(2014)]{Kartha} Kartha, S.~S., Forbes, 
D.~A., Spitler, L.~R., et al.\ 2014, \mnras, 437, 273

\bibitem[Kronberger et al.(2008)]{Kronberger} Kronberger, T., Kapferer, W., Unterguggenberger, S., Schindler, S., \& Ziegler, B.~L.\ 2008, \aap, 483, 783


\bibitem[Kruijssen et al.(2012)]{k12} Kruijssen, J.~M.~D., 
Pelupessy, F.~I., Lamers, H.~J.~G.~L.~M., et al.\ 2012, \mnras, 421, 1927 



  

  
\bibitem[Minniti(1995)]{minniti95} Minniti, D.\ 1995, \aap, 303, 468 

\bibitem[Larsen 
\& Brodie(2000)]{LB00} Larsen, S.~S., \& Brodie, J.~P.\ 2000, \aj, 120, 2938 




\bibitem[Naab et al.(2014)]{naab14} Naab, T., Oser, L., 
Emsellem, E., et al.\ 2014, \mnras, 444, 3357 

\bibitem[Norris et al.(2008)]{norris08} Norris, M.~A., Sharples, 
R.~M., Bridges, T., et al.\ 2008, \mnras, 385, 40 


\bibitem[Norris et al.(2014)]{norris14} Norris, M.~A., 
Kannappan, S.~J., Forbes, D.~A., et al.\ 2014, \mnras, 443, 1151 

\bibitem[Noordermeer et al.(2008)]{Nor08} Noordermeer, E., 
Merrifield, M.~R., Coccato, L., et al.\ 2008, \mnras, 384, 943 

\bibitem[Peng et al., 2002]{peng02} Peng, C.~Y., Ho, L.~C., Impey,
  C.~D., \& Rix, H.-W.\ 2002, \aj, 124, 266
  
\bibitem[Pota et al.(2013)]{pota13} Pota, V., Forbes, D.~A., 
Romanowsky, A.~J., et al.\ 2013, \mnras, 428, 389

\bibitem[Proctor et al.(2008)]{proctor08} Proctor, R.~N., Forbes, 
D.~A., Brodie, J.~P., \& Strader, J.\ 2008, \mnras, 385, 1709 

\bibitem[Proctor et al.(2009)]{proctor09} Proctor, R.~N., Forbes, 
D.~A., Romanowsky, A.~J., et al.\ 2009, \mnras, 398, 91 

\bibitem[Rocha et al.(2012)]{Rocha} Rocha, M., Peter, 
A.~H.~G., \& Bullock, J.\ 2012, \mnras, 425, 231 



\bibitem[Quilis et al.(2000)]{Quilis} Quilis, V., Moore, B., 
\& Bower, R.\ 2000, Science, 288, 1617 

\bibitem[Sancisi et al.(1984)]{sancisi85} Sancisi, R., van 
Woerden, H., Davies, R.~D., \& Hart, L.\ 1984, \mnras, 210, 497 


\bibitem[Shapiro et al.(2010)]{shapiro10} Shapiro, K.~L., Genzel, 
R., F{\"o}rster Schreiber, N.~M.\ 2010, \mnras, 403, L36 




\bibitem[Yoon et al.(2006)]{yoon06} Yoon, S.-J., Yi, S.~K., 
\& Lee, Y.-W.\ 2006, Science, 311, 1129 



\bibitem[Zinn(1985)]{zinn85} Zinn, R.\ 1985, \apj, 293, 424 

\bibitem[Zinn(1996)]{zinn96} Zinn, R.\ 1996, Formation of the 
Galactic Halo...Inside and Out, 92, 211 




\end{thebibliography}
\end{document}